\documentclass[aps,twocolumn,showpacs,superscriptaddress,amsmath,amssymb,nofootinbib]{revtex4-1}

\usepackage{graphicx}
\usepackage{dcolumn}
\usepackage{bm}
\usepackage{color}
\usepackage{txfonts}
\usepackage{hyperref}
\usepackage{soul}

\begin{document}
\title{Density-based and transport-based core-periphery structures in networks}

\author{Sang Hoon Lee}
\email[Corresponding author: ]{lee@maths.ox.ac.uk}
\affiliation{Oxford Centre for Industrial and Applied Mathematics, Mathematical Institute, University of Oxford, Oxford, OX2 6GG, United Kingdom}

\author{Mihai Cucuringu}
\affiliation{Program in Applied and Computational Mathematics (PACM), Princeton University, Fine Hall, 
Washington Road, Princeton, New Jersey 08544-1000, USA}
\affiliation{Department of Mathematics, University of California, Los Angeles, Los Angeles, California 90095, USA}

\author{Mason A. Porter}
\affiliation{Oxford Centre for Industrial and Applied Mathematics, Mathematical Institute, University of Oxford, OX2 6GG, United Kingdom}
\affiliation{CABDyN Complexity Centre, University of Oxford, Oxford OX1 1HP, United Kingdom}


\begin{abstract}

Networks often possess mesoscale structures, and studying them can yield insights into both structure and function.  It is most common to study community structure, but numerous other types of mesoscale structures also exist.  In this paper, we examine core-periphery structures based on both density and transport.  In such structures, core network components are well-connected both among themselves and to peripheral components, which are not well-connected to anything.  We examine core-periphery structures in a wide range of examples of transportation, social, and financial networks---including road networks in large urban areas, a rabbit warren, a dolphin social network, a European interbank network, and a migration network between counties in the United States. We illustrate that a recently developed transport-based notion of node coreness is very useful for characterizing transportation networks.  We also generalize this notion to examine core versus peripheral edges, and we show that the resulting diagnostic is also useful for transportation networks. To examine the properties of transportation networks further, we develop a family of generative models of roadlike networks.  We illustrate the effect of the dimensionality of the embedding space on transportation networks, and we demonstrate that the correlations between different measures of coreness can be very different for different types of networks.

\end{abstract}



\maketitle


\section{Introduction}
\label{sec:introduction}

Studies of networks~\cite{ComplexNetwork} initially focused on local characteristics or on macroscopic distributions (of individual nodes and edges), but it is now common to consider ``mesoscale'' structures such as communities~\cite{CommunityReview}. Indeed, there are numerous notions of community structure in networks.  For example, one can define a network's community structure based on a hard or soft partitioning of network into sets of nodes that are connected more densely among themselves than to nodes in other sets~\cite{ng2004}, and one can also examine community structure by partitioning edges~\cite{YYAhn2010}.  One can also determine community structure by taking the perspective of a dynamical system (e.g., a Markov process) on a network~\cite{Rosvall2007,jeub2013,lambiotte08}. See Ref.~\cite{CommunityReview} for myriad other notions of community structure, which have yielded insights on numerous systems in biology~\cite{anna2010,dani2011}, political science~\cite{mason2005,mucha2010}, sociology~\cite{marta2007,traud2012}, and many other areas.

Although community structure is the most widely studied mesoscale structure by far, numerous other types exist.  These include notions of role similarity~\cite{everettrole} and many types of block models~\cite{doreian}. Perhaps the most prominent block structure aside from community structure is \emph{core-periphery structure}~\cite{Borgatti1999,Holme2005,Silva2008,Rombach2012,Csermely2013}, in which connections between core nodes and other core nodes are dense, connections between core nodes and peripheral nodes are also dense (but possibly less dense than core-core connections), and peripheral nodes are sparsely connected to other nodes. Core-periphery structure provides a useful complement for community structure~\cite{Rombach2012,Csermely2013,JYang2012}.  Its origins lie in the study of social networks (e.g., in international relations)~\cite{wallerstein,Borgatti1999}, although notions such as ``nestedness'' in ecology also attempt to determine core network components~\cite{bascompte}. As with community structure, there are numerous possible ways to examine core-periphery structure, although this has seldom been explored to date.  A few different notions of core-periphery structure have been developed~\cite{Csermely2013}, although there are far fewer of these than there are notions of community structure~\cite{CommunityReview}.  

In this paper, we contrast two different notions of core-periphery structure---the block-model perspective that we discussed above and a recently-developed notion that is appropriate for transportation networks (and which need not satisfy the density properties of the block-model notion)~\cite{Cucuringu2013a}---by calculating them for several different types of empirical and computer-generated networks. Due to the rich variety of types of networks across various areas and disciplines, a wealth of different mesoscale features are possible~\cite{Onnela2012}.  We expect a block-model notion of core-periphery structure to be appropriate for social networks, whereas it can be desirable to develop transport-based notions of core-periphery structure for road networks and other transportation networks.  However, this intuition does not imply that application-blind notions cannot be useful (e.g., a recently-developed block-model notion of core-periphery structure was helpful for analyzing the London metropolitan transportation system~\cite{Rombach2012}), but it is often desirable for network notions to be driven by applications for further development.  This is also the case for community structure~\cite{CommunityReview,Onnela2012}, where measures of modularity~\cite{Newman2006}, conductance~\cite{Mahoney2009}, information cost~\cite{Rosvall2007}, and partition density (for communities of edges)~\cite{YYAhn2010} are all useful.  Core-periphery structure depends on context and application, and it is important to compare different notions of core network components when considering core agents in a social network, core banks in a financial system, core streets and intersections in a road network, and so on.

We focus on two different ways of characterizing core-periphery structures in networks: we examine density-based (or ``structural'') coreness using intuition from social networks---in which core agents either have high degree (or strength, in the case of weighted networks), are neighbors of nodes with high degree (or strength), or satisfy both properties---and we examine transport-based coreness by modifying notions of betweenness centrality~\cite{Cucuringu2013a}. To contrast these different types of core-periphery structure, we compute statistical properties of coreness measures applied to empirical networks, their correlations to each other, and their correlations to other properties of networks.  With these calculations, we obtain interesting insights on several social, financial, and transportation networks. An additional contribution of this paper is our extension of the transport-based method in Ref.~\cite{Cucuringu2013a} to allow the assignment of a coreness measure to edges (rather than just nodes).  Such a generalization is clearly important for transportation networks, for which one might want (or even need) to focus on edges rather than nodes.

The remainder of this paper is organized as follows.  In Sec.~\ref{sec:methods}, we discuss the methods that we employ in this paper for studying density-based and transport-based core-periphery structure.  We examine some social and financial networks in Sec.~\ref{sec:examples_non_transporation_networks} and several transportation networks in Sec.~\ref{sec:examples_transporation_networks}.  To illustrate the effects of spatial embedding on transportation networks, we develop a generative model of roadlike networks in Sec.~\ref{sec:examples_transporation_networks}.  We conclude in Sec.~\ref{sec:discussion}.


\section{Core-periphery structure in networks}
\label{sec:methods}


\subsection{Density-Based Core-Periphery Structure}
\label{sec:structural_CP}

Conventional definitions of core-periphery organization rely on connection densities among different sets of nodes (in the form of block models) or on structural properties such as node degree and strength.  One approach to studying core-periphery structure relies on finding a group of core nodes or assigning coreness values to nodes by optimizing an objective function~\cite{Borgatti1999,Holme2005,Silva2008,Rombach2012}.  The method introduced in Ref.~\cite{Rombach2012}, which generalizes the basic (and best known) formulation in~\cite{Borgatti1999}, is particularly flexible.  For example, one can detect distinct cores in a network, and one can consider either discrete or continuous measures of coreness.
This notion was used recently to examine the roles of brain regions for learning a simple motor task in functional brain networks~\cite{Bassett2013}. 

In the method of Ref.~\cite{Rombach2012}, one seeks to calculate a centrality measure of coreness called a ``core score'' (CS) using the adjacency-matrix elements $\{ W_{ij} \}$, where $i,j \in \{1,\ldots N\}$, the network has $N$ nodes, and the value $W_{ij}$ indicates the weight of the connection between nodes $i$ and $j$.  For directed networks (see the discussion below), we use $W_{ij}$ to denote the weight of the connection from node $i$ to node $j$.  When $W_{ij} = 0$, there is no edge between $i$ and $j$. We insert the core-matrix elements $\{ C_{ij} \}$ into the core quality
\begin{equation}
	R({\alpha,\beta}) = \sum_{i,j} W_{ij} C_{ij}(\alpha,\beta)\,, \label{core_quality}
\end{equation}
where the parameter $\alpha \in [0,1]$ determines the sharpness of the core-periphery division and $\beta \in [0,1]$ determines the fraction of
core nodes. We decompose the core-matrix elements into a product form, $C_{ij}(\alpha,\beta) = C_i(\alpha,\beta) C_j(\alpha,\beta)$, where
\begin{equation}
	C_i (\alpha,\beta) = \begin{cases}
\frac{\displaystyle i(1-\alpha)}{\displaystyle 2 \lfloor \beta N \rfloor} , & i \le \lfloor \beta N \rfloor , \\
\frac{\displaystyle (i-\lfloor \beta N \rfloor)(1-\alpha)}{\displaystyle 2(N-\lfloor \beta N \rfloor)} + \frac{\displaystyle 1+\alpha}{\displaystyle 2} , & i > \lfloor \beta N \rfloor 
\end{cases}
\label{transition_function}
\end{equation}
are the elements of a core vector. Reference~\cite{Rombach2012} also discusses the use of alternative ``transition functions'' to the one in Eq.~(\ref{transition_function}).

We wish to determine the core-vector elements in (\ref{transition_function}) so that the core quality in Eq.~(\ref{core_quality}) is maximized.  This yields a CS for node $i$ of
\begin{equation}
	\textrm{CS}(i) = Z \sum_{(\alpha,\beta)} C_i (\alpha,\beta) R(\alpha,\beta)\,,
\label{CS_formula}
\end{equation}
where the normalization factor $Z$ is determined so that the maximum value of CS over the entire set of nodes is 1. In practice, we perform the optimization using some computational heuristic and some sample of points in the parameter space with coordinates $(\alpha,\beta) \in [0,1] \times [0,1]$.  As in Ref.~\cite{Rombach2012}, we use simulated annealing~\cite{Kirkpatrick1983} (with the same cooling schedule as in that paper).  This adds stochasticity to the method. The core-quality landscape tends to be less sensitive to $\alpha$ than it is to $\beta$, so one can reduce the number of $\alpha$ values for computationally expensive situations if it is necessary.  For all examples in this paper, we use the sampling resolutions $\Delta \alpha = \Delta \beta = 0.01$ and thus consider $101^2$ evenly-spaced points in the ($\alpha,\beta)$ plane.

For directed networks, one can still technically compute CS values because Eq.~\eqref{core_quality} is still valid when the matrix ${\bf W}$ is asymmetric, so that is what we will do in the present paper.  However, it seems strange to produce only one set of core scores rather than two sets of them (just like one wishes to compute both in-degrees from out-degrees in a directed network), and the $i \to j$ and $j \to i$ interactions are confounded in Eq.~\eqref{core_quality} because $W_{ij}$ and $W_{ji}$ appear on equal footing.  (The transport-based notions of core-periphery structure that we will discuss in Sec.~\ref{sec:transport_CP} apply naturally to both directed and undirected networks; this follows the spirit of directed flow on networks.)  It is both interesting and desirable to investigate density-based notions (e.g., via block models) of core-periphery structure for directed networks, but we will not pursue that in this paper.  Such notions would allow one to distinguish between core sources and core sinks.


\subsection{Transport-Based Core-Periphery Structures}
\label{sec:transport_CP}

Notions of betweenness centrality (BC) are useful for characterizing transportation properties of networks~\cite{ComplexNetwork,Freeman1977,KIGoh2001}, and ideas based on short paths have been used to examine core-periphery structure~\cite{Silva2008,rwcore,Cucuringu2013a}.  

In our discussion of transport-based core-periphery structures, we will draw on a notion that was introduced in Ref.~\cite{Cucuringu2013a} and was inspired by geodesic node betweenness centrality. In this paper, we will also define an analogous notion for core and peripheral edges. The basic idea is that core network components (e.g., nodes or edges) are used more frequently for transportation, as quantified by a BC or a similar diagnostic, than peripheral components.  To amplify the usage of connections from arbitrary parts of a network to core parts, we consider ``backup paths,'' which are the shortest paths that remain after some part or parts of a network have been removed, .

We consider networks that can be either weighted or unweighted and either directed or undirected. Let the set of edges be denoted by $\mathbb{E} = \{ (j,k) | $ where node $j$ is connected to $k \}$. The ``path score'' (PS) for node $i$ is a notion of centrality and is defined by~\cite{Cucuringu2013a}
\begin{equation}
	\textrm{PS}(i) = \frac{\displaystyle 1}{\displaystyle | \mathbb{E} |} \sum_{(j,k) \in \mathbb{E}} \sum_{\{ p_{jk} \}} \sigma_{jik} [ \mathbb{E} \setminus (j,k) ]\,,
\label{PS_formula}
\end{equation}
where $\sigma_{jik} [ \mathbb{E} \setminus (j,k) ] = 1/|\{ p_{jk} \}|$ if node $i$ is in the set $\{ p_{jk} \}$ that consists of ``optimal backup paths'' from node $j$ to node $k$, where we stress that {\em the edge} $(j,k)$ {\em is removed from} $\mathbb{E}$, and $\sigma_{jik} [ \mathbb{E} \setminus (j,k) ] = 0$ otherwise. 

Just as betweenness centralities can be defined for edges~\cite{Girvan2002} (as well as other network components) in addition to nodes, it is useful to calculate a PS for edges. We define the path score $\textrm{PS}(l)$ of edge $l$ similarly to the node PS from Eq.~(\ref{PS_formula}), except that we replace the node $i$ with the edge $l$. That is, $\sigma_{jlk} = 1/|\{ p_{jk} \}|$ if $l$ is a part of one of the optimal backup paths from node $j$ to node $k$; otherwise, $\sigma_{jlk} = 0$. Calculating a value of coreness for edges is particularly relevant for networks in which edges are fundamentally important physical, logical, or social entities~\cite{YYAhn2010}. 

A PS for a network component quantifies its importance by examining centrality scores after other components have been removed.  The importance of edges in road networks has been studied previously using the different (and much more computationally demanding) task of quantifying the importance of a removed edge by calculating BCs both before and after its removal~\cite{marc2012}.  Backup paths have also been studied in the context of percolation~\cite{eduardo}.

The notion of a PS is deterministic whenever it is based on a deterministic notion of betweenness.  (Recall that the use of simulated annealing as a computational heuristic to calculate CSs is a source of stochasticity for the formulation of core-periphery structure in Sec.~\ref{sec:structural_CP}.) Even when there exists more than one optimal backup path (which, in practice, occurs mostly for unweighted networks), all of the optimal paths $\{ p_{jk} \}$ contribute equally to the PS.  One can, of course, incorporate stochasticity by constructing a PS based on a stochastic notion of centrality (e.g., random-walk node betweenness~\cite{rwbetween}).

In the present paper, we calculate PS values based on shortest paths (i.e., ``geodesic PS values'') as well as PS values based on greedy-spatial-navigation (GSN) paths, which are constructed from local directional information and correspond to a more realistic form of navigation than geodesic paths for spatially embedded networks~\cite{SHLee2012}.  We use the acronym PS to indicate a path score that is determined via shortest paths and GSNP for a path score that is determined via a GSN path, which we define as follows~\cite{GSN_details}. Consider a network with $N$ nodes that is embedded in $\mathbb{R}^d$, and suppose that the coordinates of the nodes are $\{\mathbf{r}_1,\dots,\mathbf{r}_N\}=\{[x_1(1),x_2(1),\ldots,x_d(1)],\ldots ,[x_1(N),x_2(N),\ldots,x_d(N)]\}$.
Assume that an agent stands at a node $i$ and wishes to 
travel to node $t$. Let $\mathbf{v}_{i,j}=\mathbf{r}_j-\mathbf{r}_i$ be the vector from node $i$ to node $j$, and let
$\theta_j = \cos^{-1} [\mathbf{v}_{i,t} \cdot \mathbf{v}_{i,j} / (|\mathbf{v}_{i,t}||\mathbf{v}_{i,j}|)]$ be 
the angle between $\mathbf{v}_{i,t}$ and $\mathbf{v}_{i,j}$. A greedy navigator considers the set $\Gamma(i)$ of neighbors of $i$ and it moves to the neighbor $j \in \Gamma(i)$ that has the smallest $\theta_j$, where ties are broken by taking a neighbor uniformly at random among the neighbors with the smallest angle. 
If all neighbors $j \in \Gamma(i)$ have been visited, then the navigator 
goes back to the node that it left to reach $i$.
This procedure is repeated until node $t$ is reached (which will happen eventually if $G$ is connected, or, more generally, if $i$ and $j$ belong to the same component).

It is important at this stage to comment about weight versus ``distance'' in weighted networks. In a weighted network, a larger weight represents a closer or stronger relation. If we are given such a network (with weight-matrix elements $W_{ij}$), we construct a distance matrix whose elements are $D_{ij} = 1/W_{ij}$ for nonzero $W_{ij}$ and $D_{ij} = 0$ when $W_{ij} = 0$.  We then use the distance matrix to determine the length of a path and in all of our calculations of PSs, GSNPs, and BCs.  Alternatively, we might start with a set of network distances or Euclidean distances, and then we can use that information directly.  In this paper, we will consider transportation networks that are embedded in $\mathbb{R}^2$ and $\mathbb{R}^3$. In contrast to CSs, we can calculate PSs for directed networks very naturally simply by restricting ourselves to directed paths.


\section{Social and financial networks}
\label{sec:examples_non_transporation_networks}

We now examine some social and financial networks, as it is often argued that such networks possess a core-periphery structure.  Indeed, the intuition behind density-based core-periphery structure was developed from studies of social networks~\cite{wallerstein,Borgatti1999,Rombach2012}.

As discussed in Sec.~\ref{sec:transport_CP}, we highlight an important point for weighted networks.  In such networks, each edge has a value associated with it. We consider data associated with such values that come in one of two forms.  In one form, we have a matrix entry $W_{ij}$ for which a larger value indicates a closer (or stronger) relationship between nodes $i$ and $j$ (where $i \neq j$).  In this case, we have a weighted adjacency matrix ${\bf W}$ whose elements are $W_{ij}$.  In the second form, we have a matrix entry $D_{ij}$ for which a larger value indicates a more distant (literally, in the case of transportation networks) or weaker relationship between nodes $i$ and $j$ (where $i \neq j$).  In this case, the elements $D_{ij}$ yield a distance matrix ${\bf D}$, and we calculate weighted adjacency matrix elements using the formula $W_{ij} = 1/D_{ij}$ (for $i \ne j$) and $D_{ij} = 0$ when $W_{ij} = 0$.


\subsection{Dolphin Social Network}
\label{sec:dolphin_social_network}

\begin{table*}[t]
\caption{Pearson and Spearman correlation coefficients between various pairs of core and centrality values (CS, PS, and BC) for some social and financial networks.  We calculated these correlations using the \textsc{SciPy} package in \textsc{Python}~\cite{SciPy}.  In parentheses, we give two-tailed $p$-values with the null hypothesis of absence of correlation. We use the $^{\dagger}$ symbol when all BC values are the same; in this case, it is not meaningful to compute the correlations.
}
\label{SummaryTable_social1}
\begin{tabular}{llllll}
\hline\hline
Network & Correlation & CS vs PS & CS vs BC & PS vs BC & PS vs BC \\
 & & (nodes) & (nodes) & (nodes) & (edges) \\
\hline
Dolphin~\cite{Lusseau2003} & Pearson & $0.811$ & $0.426$ & $0.452$ & $-0.249$ \\
 & & $(1.26 \times 10^{-15})$ & $(5.64 \times 10^{-4})$ & $(2.28 \times 10^{-4})$ & $(1.54 \times 10^{-3})$ \\
\cline{2-6}
 & Spearman & $0.835$ & $0.704$ & $0.715$ & $-0.418$ \\
 & & $(3.16 \times 10^{-17})$ & $(1.79 \times 10^{-10})$ & $(6.88 \times 10^{-11})$ & $(4.19 \times 10^{-8})$ \\
\hline
Stock~\cite{YahooFinance} & Pearson & $0.222$ & ${\dagger}$ & ${\dagger}$ & ${\dagger}$ \\
 & & $(4.61 \times 10^{-7})$ & ${\dagger}$ & ${\dagger}$ & ${\dagger}$ \\
\cline{2-6}
 & Spearman & $0.130$ & ${\dagger}$ & ${\dagger}$ & ${\dagger}$ \\
 & & $(3.53 \times 10^{-3})$ & ${\dagger}$ & ${\dagger}$ & ${\dagger}$ \\
\hline
\hline
\end{tabular}
\end{table*}

\begin{figure*}
\begin{tabular}{ll}
(a) & (b) \\
\hspace{.2in}
\includegraphics[width=0.9\columnwidth]{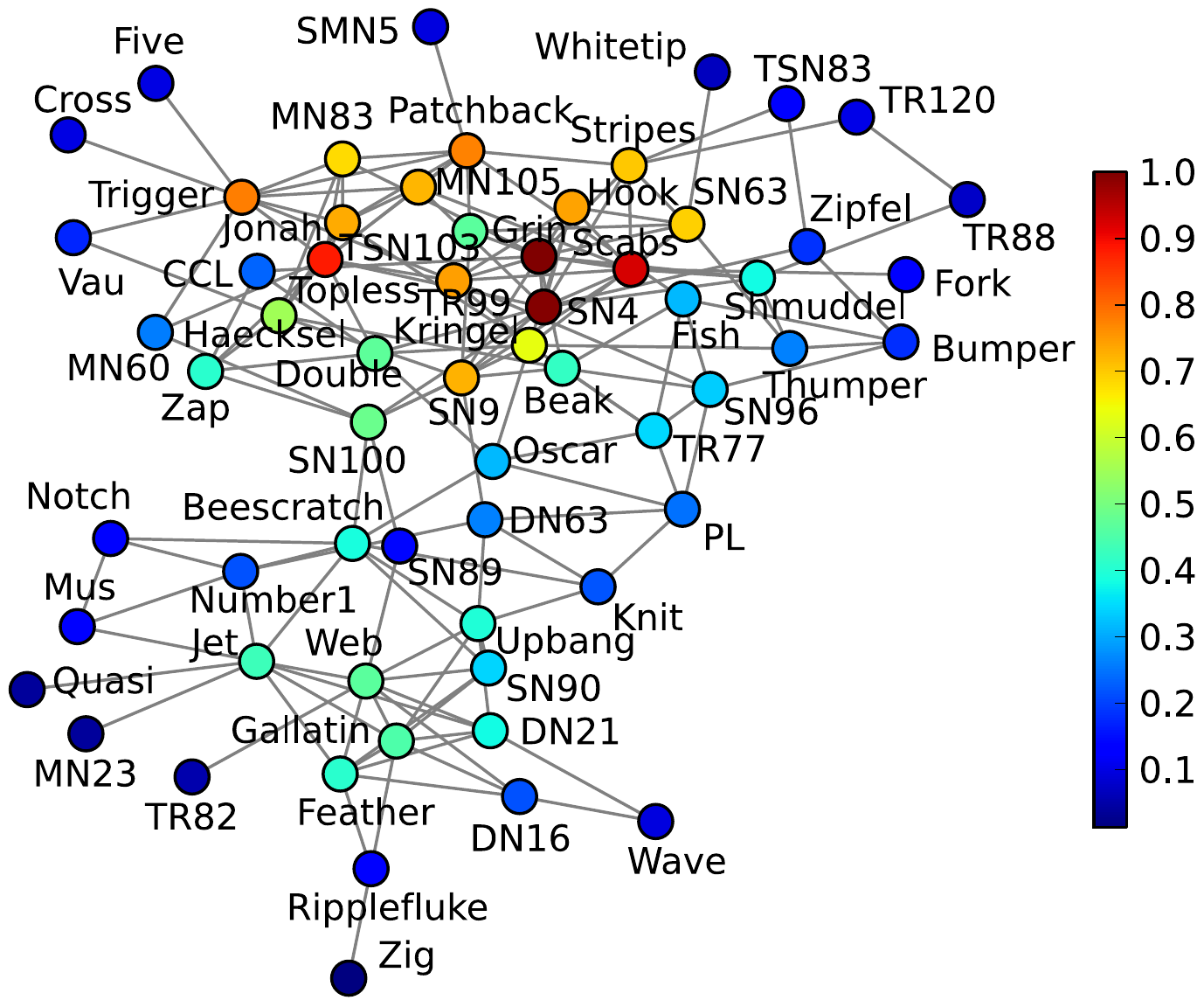} &
\hspace{.2in}
\includegraphics[width=0.9\columnwidth]{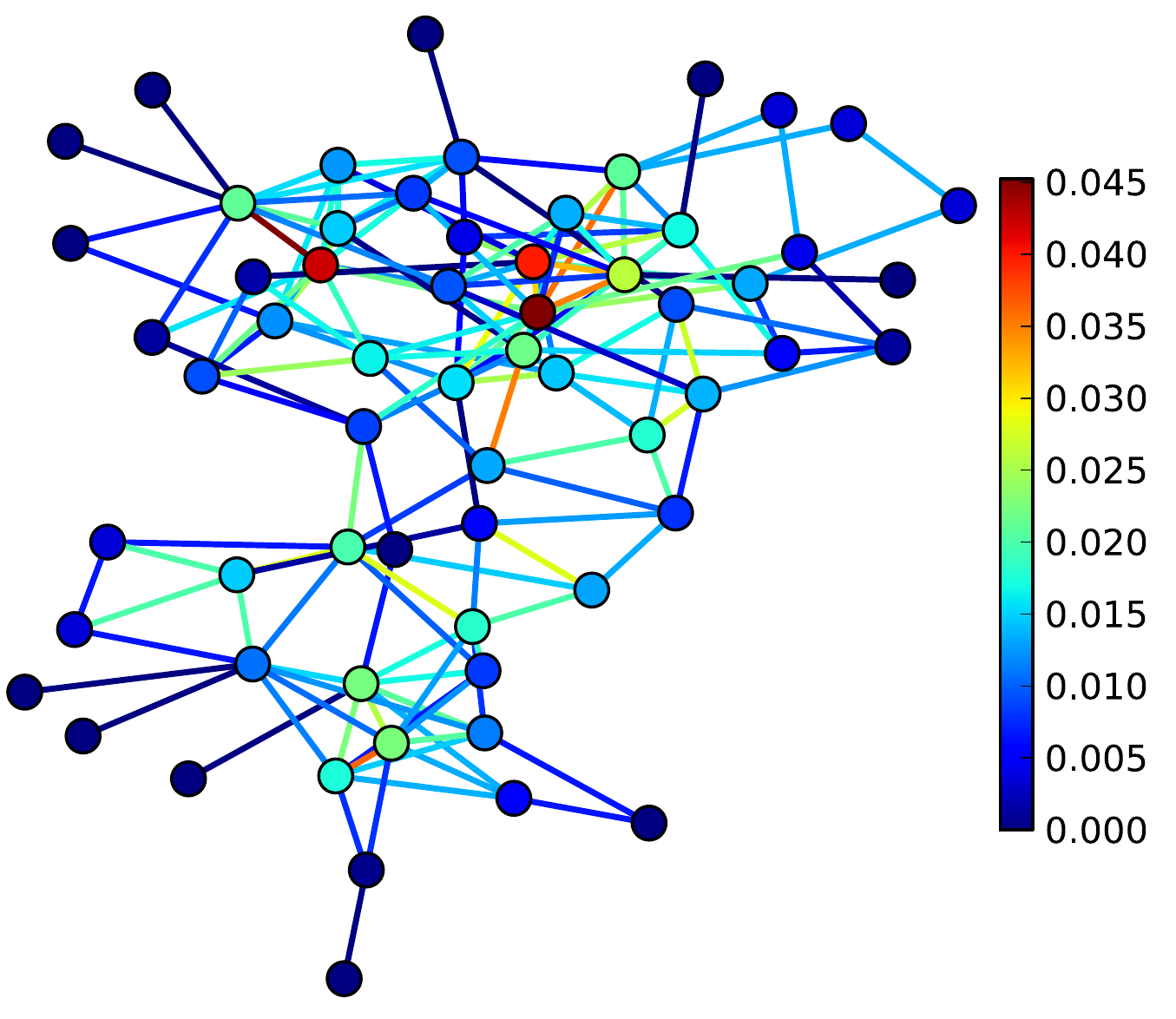} 
\end{tabular}
\caption{The dolphin social network, for which we color (a) the nodes using CS values and (b) the nodes and edges using, respectively, node and edge PS values. The same color bar in (b) applies to both node and edge PSs. We show the dolphins' names  in (a).  In both panels, we position the nodes using a Kamada-Kawai force-directed graph drawing algorithm~\cite{Kamada1989}, for which we used the ``\textsc{graphviz\_layout}'' function in the \textsc{NetworkX} package for \textsc{Python}~\cite{graphviz}.
}
\label{dolphin_network_with_CS_PS}
\end{figure*}

As a small example to set the stage, we consider the (unweighted and undirected) social network between $62$ bottlenose dolphins ({\it Tursiops} spp.) in a community living near Doubtful Sound, New Zealand~\cite{Lusseau2003,NewmanWebsite}. In Fig.~\ref{dolphin_network_with_CS_PS}, we color this network using the CS values of nodes and the geodesic PS values of nodes and edges.  Geodesic node betweenness was used previously to examine important dolphins in this network, and examining coreness measures allows one to build on such insights. The five dolphins with the largest geodesic BC values are (in order) SN100, Beescratch, SN9, SN4, and DN63~\cite{Lusseau2004}; the five dolphins with the largest CS values are (in order) Grin, SN4, Scabs, Topless, and Trigger; and the five dolphins with the largest PS values are (in order) SN4, Topless, Grin, Scabs, and Gallatin.  Some dolphins seem to be important according to all of these measures, but other names change.

As shown in Table~\ref{SummaryTable_social1}, the two coreness measures
(CS and PS) are correlated with each other much more strongly than either of them is correlated with BC. The coreness measures that we employ can be used to further investigate the dolphins' social roles (some of which have been described previously~\cite{Lusseau2003,Lusseau2004,Lusseau2007}).
For instance, dolphins that exhibit side flopping (SF) or upside-down lobtailing (ULT) behaviors~\cite{Lusseau2007} have a wide range of coreness values, so such behaviors do not seem to relate to whether a dolphin is a core node. As SF and ULT behaviors are known to play communication roles~\cite{Lusseau2006}, this might illustrate that communication is necessary throughout the social hierarchy of dolphins rather than only occurring in specific levels of it.

We also identify the edges with the largest PS values.  In order, these edges\footnote{For example like this one, we are using brackets rather than parentheses to indicate the edges because it is easier to read.} correspond to the dolphin pair [Topless, Trigger], [Feather, Gallatin], [Stripes, SN4], [SN4, Scabs], and [Kringel, Oscar]. The edges with the largest geodesic BC values are (in order) [Beescratch, SN100], [SN9, DN63], [Jet, Beescratch], [SN100, SN4], and [SN89, SN100]. As shown in Fig.~\ref{dolphin_network_with_CS_PS}, ``bridge'' edges such as [Beescratch, SN100] or [SN9, DN63] that connect two large communities have the largest BC values. Naturally, these edges are not core edges. Indeed, as shown in Table~\ref{SummaryTable_social1}, geodesic edge PS and geodesic edge BC are {\em negatively} correlated.


\subsection{Interbank Network}
\label{sec:interbank_network}

\begin{figure*}
\begin{tabular}{ll}
(a) & (b) \\
\hspace{.2in}
\includegraphics[height=0.75\columnwidth]{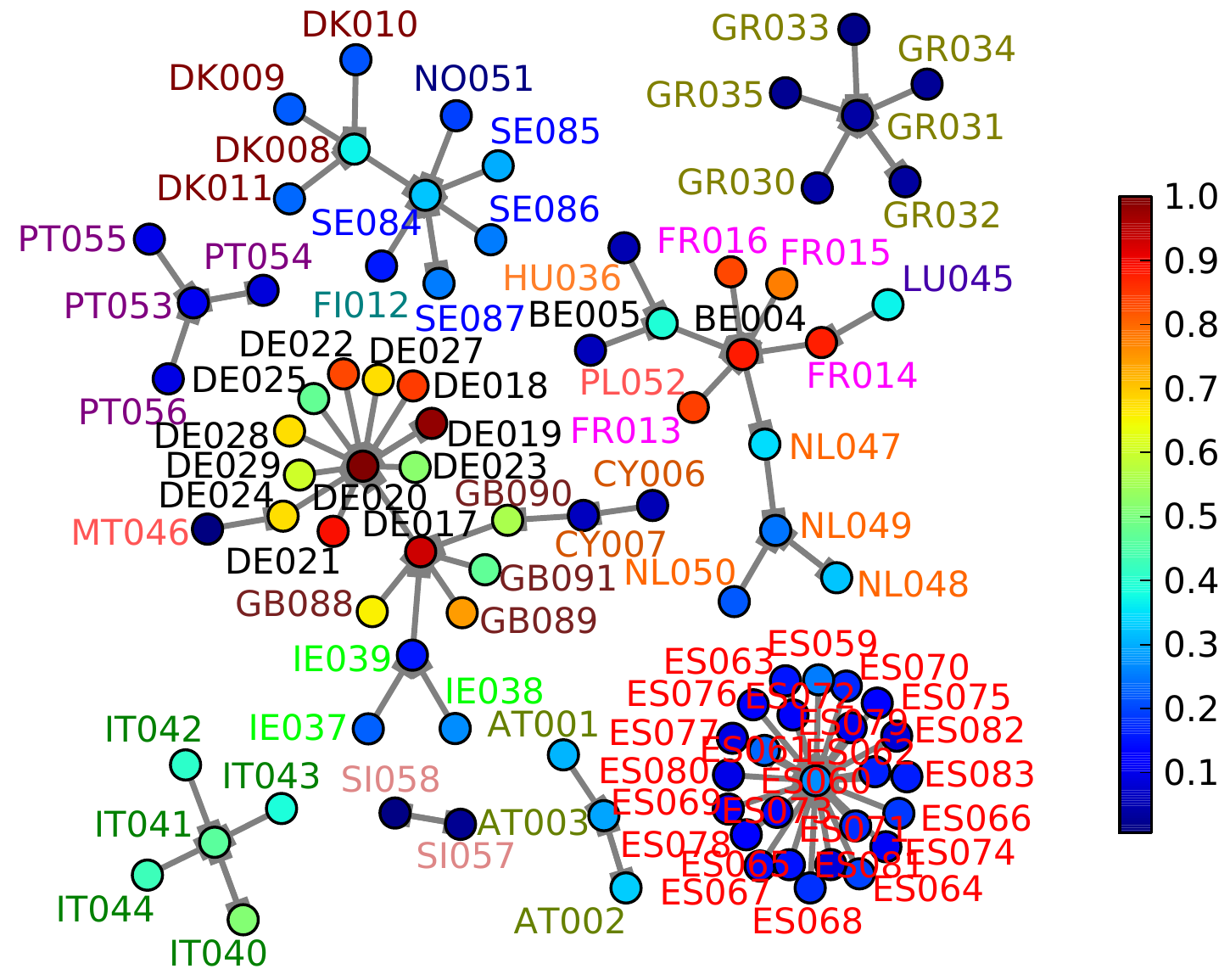} &
\hspace{.2in}
\includegraphics[height=0.75\columnwidth]{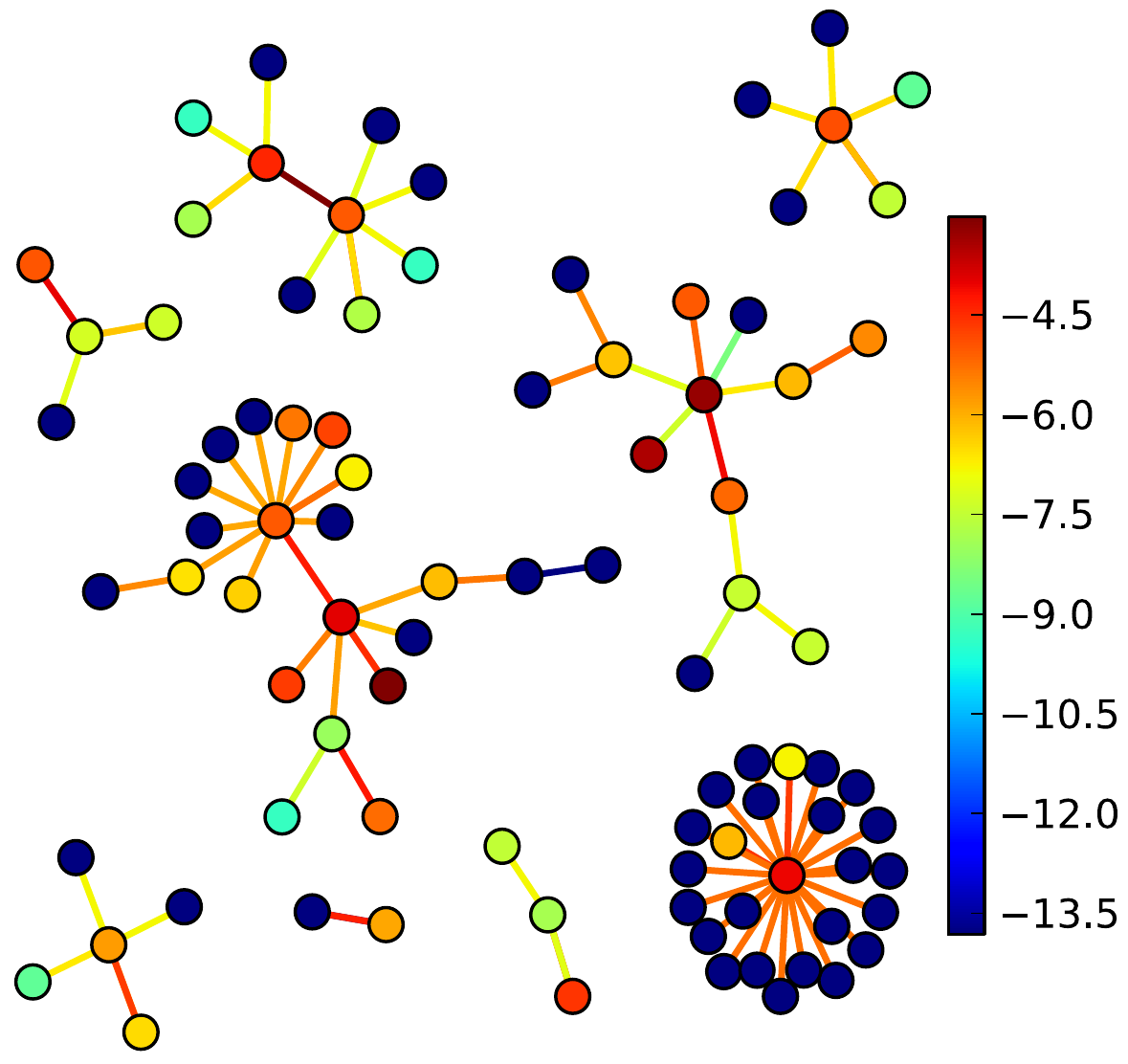} 
\end{tabular}
\caption{The maximum relatedness subnetwork (MRS)~\cite{GooglingSocialInteractions} of a European interbank network. We color the nodes and edges based on (a) CS and (b) $\ln(\textrm{PS} + 10^{-6})$ computed using the original (much denser) network.  (We use a logarithm for visualization because of the heterogeneity of the PS values.) In (a), we identify the directions of edges using thick stubs to indicate arrowheads~\cite{NetworkXEdges}.  We also give the banks' codes in (a).  The corresponding bank identities are listed in Ref.~\cite{InterbankNetwork}.  Different font colors represent different countries. In both panels, we position the nodes using a Kamada-Kawai force-directed graph drawing algorithm~\cite{Kamada1989}, for which we used the \textsc{graphviz\_layout} function in the \textsc{NetworkX} package for \textsc{Python}~\cite{graphviz}.
}
\label{interbank_network_with_CS_PS}
\end{figure*}

\begin{table*}[t]
\caption{Pearson and Spearman correlation values between various core and centrality values (CS, PS, and BC) for several social and financial networks. In parentheses, we give two-tailed $p$-values with the null hypothesis of absence of correlation. For the interbank network, we also show the correlation between the measures and the banks' tier-1 capital (which corresponds to their ``size''~\cite{InterbankNetwork,Tier1Capital}). For the U.S. migration network, we also show correlations between population of counties (which corresponds to their size) versus other measures. 
When we write that a $p$-value is $0.0$, it means that this value is smaller than the minimum (approximately $2.23 \times 10^{-308}$) of the floating-point variables in \textsc{Python}.
We use the same \textsc{SciPy} package in \textsc{Python}~\cite{SciPy} as in Table~\ref{SummaryTable_social1}.
}
\label{SummaryTable_social2}
\begin{ruledtabular}
\begin{tabular}{lllllllll}
 Network & Correlation & CS vs PS & CS vs BC & PS vs BC & PS vs BC & size vs CS & size vs PS & size vs BC \\
 & & (nodes) & (nodes) & (nodes) & (edges) & (nodes) & (nodes) & (nodes) \\
\hline
Interbank~\cite{InterbankNetwork} & Pearson & $0.430$ & $0.410$ & $0.929$ & $0.885$ & $0.533$ & $0.605$ & $0.685$ \\
& & $(2.39 \times 10^{-5})$ & $(6.09 \times 10^{-5})$ & $(6.98 \times 10^{-40})$ & $(0.0)$ & $(6.33 \times 10^{-8})$ & $(2.65 \times 10^{-10})$ & $(9.44 \times 10^{-14})$ \\
\cline{2-9}
 & Spearman & $0.499$ & $0.502$ & $0.873$ & $0.689$ & $0.667$ & $0.610$ & $0.611$ \\
 & & $(5.47 \times 10^{-7})$ & $(4.70 \times 10^{-7})$ & $(3.02 \times 10^{-29})$ & $(0.0)$ & $(6.97 \times 10^{-13})$ & $(1.75 \times 10^{-10})$ & $(1.56 \times 10^{-10})$ \\
\hline
US Migration & Pearson & $0.537$ & $0.515$ & $0.840$ & $0.860$ & $0.408$ & $0.479$ & $0.241$ \\
${\bf W}$~\cite{MigrationCensus} & & $(5.11 \times 10^{-229})$ & $(2.27 \times 10^{-208})$ & $(0.0)$ & $(0.0)$ 
& $(9.73 \times 10^{-124})$ & $(9.13 \times 10^{-176})$ & $(9.48 \times 10^{-42})$ \\
\cline{2-9}
 & Spearman & $0.579$ & $0.510$ & $0.893$ & $0.898$ & $0.667$ & $0.557$ & $0.328$ \\
 & & $(2.28 \times 10^{-274})$ & $(1.72 \times 10^{-203})$ & $(0.0)$ & $(0.0)$ & $(0.0)$ & $(1.42 \times 10^{-250})$ & $(5.27 \times 10^{-78})$ \\
\hline
US Migration & Pearson & $0.196$ & $0.183$ & $0.988$ & $0.962$ & $0.444$ & $0.828$ & $0.800$ \\
${\bf W^{\textrm{raw}}}$~\cite{MigrationCensus} & & $(6.10 \times 10^{-28})$ & $(1.52 \times 10^{-24})$ & $(0.0)$ & $(8.42 \times 10^{-5})$ & $(1.18 \times 10^{-148})$ & $(0.0)$ & $(0.0)$ \\
\cline{2-9}
 & Spearman & $0.698$ & $0.621$ & $0.942$ & $0.832$ & $0.977$ & $0.710$ & $0.635$ \\
 & & $(0.0)$ & $(0.0)$ & $(0.0)$ & $(5.77 \times 10^{-106})$ & $(0.0)$ & $(0.0)$ & $(0.0)$ \\
\end{tabular}
\end{ruledtabular}
\end{table*}

It has been argued that many financial systems exhibit core-periphery structures~\cite{Haldane2011,Lloyd2013}, but few scholars have complemented such claims with quantitative 
calculations of such structures.  A couple of notable exceptions include Refs.~\cite{CraigWorkingPaper,LuxWorkingPaper}, which used a method based on that in Ref.~\cite{Borgatti1999}. In this section, we examine core-periphery structure in an interbank credit exposure network. The nodes are banks, and a weighted and directed edge indicates an exposure from a lending bank to a borrowing bank. The magnitude of (credit) exposure indicates the extent to which the lender is exposed to the risk of loss in the event of the borrower's default~\cite{CreditExposure}.
We use data from the European Banking Authority~\cite{EBA} report on interbank exposures. It considered $90$ medium-to-large European banks~\cite{InterbankNetwork,FinancialStabilityReview}.
In principle, there is a directed and weighted edge $W_{b'b}^{\textrm{real}}$  between the lending bank $b$ and the borrowing bank $b'$.  However, data is only available for the country $c$ of a bank $b$.  This yields a matrix with components
\begin{equation*}
	E_{cb} = \sum_{b' \in \mathsf{C}(c)} W_{b'b}^{\textrm{real}} \,,
\end{equation*}	
where the set $\mathsf{C}(c)$ consists of the banks that belong to country $c$. Therefore, we assume that each $E_{cb}$ value is distributed equally among all banks $b' \neq b$ in country $c$ (i.e., except for the lending bank).  This yields an approximate weight between $b'$ and $b$ of
\begin{equation*}
	W_{b'b} = \frac{E_{cb}}{\left|\mathsf{C}(c) \setminus\{b\}\right|}\,,
\end{equation*}
and the associated distance-matrix element is $D_{b'b} = 1 / W_{b'b}$.
(We only consider node pairs with $E_{cb} \neq 0$ as edges, so $W_{b'b} \neq 0$.) Of course, one can distribute $E_{bc}$ in other ways, but we choose to use the equal-distribution scheme in the absence of additional information.
One obtains a different network with other choices, which can (of course) affect core-periphery structures.  Our choice in this paper corresponds to the one that European Banking Authority
made for their risk analysis~\cite{EBA,InterbankNetwork}. We use this example to illustrate core-periphery structures in financial 
systems~\cite{Haldane2011,Lloyd2013,CraigWorkingPaper,LuxWorkingPaper}.

The interbank credit exposure network is dense, so it is hard to visualize the core scores directly.
Therefore, after calculating the CS and PS values from the original network, we visualize the values overlaid on its maximum relatedness
subnetwork (MRS)~\cite{GooglingSocialInteractions}. An MRS is a subnetwork that is constructed as follows: for each node, we examine the weight of each of its edges and keep only the single directed edge with maximum weight.  (When there are ties, we keep all of the edges with the maximum weight.) In Fig.~\ref{interbank_network_with_CS_PS}, we show the CS values of the original weighted network (with adjacency matrix elements $W_{b'b}$) and the node PS values (with optimal paths that minimize the sum of the reciprocals of the weights) of the interbank network.  Only a few very large PS values dominate the system.  In order, these are HSBC Holdings plc (UK: GB089), Dexia (Belgium: BE004), BNP Paribas (France: FR013), Deutsche Bank (Germany: DE017), and Banco Bilbao Vizcaya Argentaria (Spain: ES060). 

In addition to the core-periphery structure, the MRS visualization illustrates that a bank's country
is crucial for the organization of its ``backbone'' structure. The banks are well-clustered according to their countries, and a few banks play the role of ``broker'' banks across different countries. The broker banks include the Nordic cluster (with Swedish, Danish, Norwegian,
and Finnish banks), the Germany-U.K.-Ireland cluster, and the France-Belgium-Netherlands-Luxembourg-Hungary-Poland cluster. In contrast to the dolphin social network that we examined in Sec.~\ref{sec:dolphin_social_network}, the CS and geodesic node PS values are less or comparably correlated to each other than either quantity is to geodesic BC (see Tables~\ref{SummaryTable_social1} and~\ref{SummaryTable_social2}).
However, the banks' tier-1 capital~\cite{Tier1Capital} is similarly correlated to each of the CS, PS, and BC values (see Table~\ref{SummaryTable_social2}).


\subsection{Stock-Market Correlation Network}
\label{sec:SP500_network}

\begin{table*}[t]
\caption{Top core nodes in the S\&P 500 and exchange-traded funds (ETFs) correlation network. We show the rank ordering based on both CS and PS values, and we mark ETFs with a ${\ddagger}$ symbol.}
\label{SP500_ETFs_Top20}
\begin{ruledtabular}
\begin{tabular}{rrr}
Rank & CS (value) & PS (value) \\
\hline
1 & Vanguard Large-Cap Index Fund$^{\ddagger}$ $(1.000)$ & Guggenheim S\&P 500 Equal Weight$^{\ddagger}$ $(3.42 \times 10^{-1})$ \\
2 & Guggenheim S\&P 500 Equal Weight$^{\ddagger}$ $(0.999)$ & iShares Russell 1000 Index Fund$^{\ddagger}$ $(9.52 \times 10^{-2})$ \\
3 & iShares Russell 1000 Index Fund$^{\ddagger}$ $(0.992)$ & Vanguard Large-Cap Index Fund$^{\ddagger}$ $(8.90 \times 10^{-2})$ \\
4 & iShares Core S\&P 500 ETF$^{\ddagger}$ $(0.990)$ & iShares Core S\&P 500 ETF$^{\ddagger}$ $(8.82 \times 10^{-2})$ \\
5 & SPDR S\&P 500 ETF$^{\ddagger}$ $(0.982)$ & Consumer Discret Select Sector SPDR$^{\ddagger}$ $(7.72 \times 10^{-2})$ \\
6 & iShares S\&P 100 Index Fund$^{\ddagger}$ $(0.979)$ & Financial Select Sector SPDR$^{\ddagger}$ $(4.80 \times 10^{-2})$ \\
7 & iShares Morningstar Large Core Index Fund$^{\ddagger}$ $(0.978)$ & Energy Select Sector SPDR$^{\ddagger}$ $(3.91 \times 10^{-2})$ \\
8 & First Trust Large Cap Core AlphaDEX Fund$^{\ddagger}$ $(0.978)$ & SPDR S\&P 500 ETF$^{\ddagger}$ $(3.77 \times 10^{-2})$ \\
9 & Vanguard Mega Cap ETF$^{\ddagger}$ $(0.971)$ & Utilities Select Sector SPDR$^{\ddagger}$ $(3.35 \times 10^{-2})$ \\
10 & RevenueShares Large Cap Fund$^{\ddagger}$ $(0.968)$ & Industrial Select Sector SPDR$^{\ddagger}$ $(3.00 \times 10^{-2})$ \\
11 & Consumer Discret Select Sector SPDR$^{\ddagger}$ $(0.967)$ & Health Care Select Sector SPDR$^{\ddagger}$ $(2.50 \times 10^{-2})$ \\
12 & Industrial Select Sector SPDR$^{\ddagger}$ $(0.963)$ & Consumer Staples Select Sector SPDR$^{\ddagger}$ $(2.46 \times 10^{-2})$ \\
13 & Financial Select Sector SPDR$^{\ddagger}$ $(0.961)$ & Technology Select Sector SPDR$^{\ddagger}$ $(2.33 \times 10^{-2})$ \\
14 & Guggenheim Russell Top 50 ETF$^{\ddagger}$ $(0.957)$ & Technology Select Sector SPDR$^{\ddagger}$ $(2.08 \times 10^{-2})$ \\
15 & PowerShares Value Line Timeliness Select Portfolio$^{\ddagger}$ $(0.955)$ & iShares S\&P 100 Index Fund$^{\ddagger}$ $(1.60 \times 10^{-2})$ \\
16 & Technology Select Sector SPDR$^{\ddagger}$ $(0.951)$ & iShares Morningstar Large Core Index Fund$^{\ddagger}$ $(2.43 \times 10^{-3})$ \\
17 & Technology Select Sector SPDR$^{\ddagger}$ $(0.950)$ & Vanguard Mega Cap ETF$^{\ddagger}$ $(2.11 \times 10^{-3})$ \\
18 & iShares KLD Select Social Index Fund$^{\ddagger}$ $(0.947)$ & Guggenheim Russell Top 50 ETF$^{\ddagger}$ $(2.06 \times 10^{-3})$ \\
19 & Energy Select Sector SPDR$^{\ddagger}$ $(0.947)$ & First Trust Large Cap Core AlphaDEX Fund$^{\ddagger}$ $(2.00 \times 10^{-3})$ \\
20 & Invesco Ltd. $(0.932)$ & Vornado Realty Trust $(9.86 \times 10^{-4})$ \\
\end{tabular}
\end{ruledtabular}
\end{table*}

As a second example of a financial network, we consider a complete, undirected, weighted stock-market network that consists of Standard and Poor (S\&P) 500 constituents along with some index exchange-traded funds (ETFs).  A weighted edge exists between every pair of nodes based on the pairwise similarities of their times series. We downloaded the (time-dependent) prices of S\&P 500 constituents and index ETFs from the Yahoo!\,Finance website~\cite{YahooFinance}. Our selection criterion was that an index or ETF time series should contain at least 1000 time points of daily prices (4 September 2009--26 August 2013).  This yields a data set that consists of time series for 9 ETFs corresponding to the sector divisions listed in Ref.~\cite{SPDR}, their component companies (of which there are 478 in total), and 17 large-cap blend equities ETFs in Ref.~\cite{SP500_ETF_list}.  For each of the 504 total time series, we calculate the daily log return: $\ln\{[\textrm{closing price }(t)]/[\textrm{opening price }(t)]\}$ on day $t$~\cite{MantegnaStanleyBook}.  To obtain the edge weights $W_{ij}$ and distances $D_{ij} = 1/W_{ij}$ in our network, we calculate the Pearson correlation coefficient $r_{ij}$ (which we subsequently shift) between each pair of daily log return series.  Specifically, $W_{ij} = (1 + r_{ij})/2$ (for $i \neq j$), where $W_{ij}$ is the weight of the edge between nodes $i$ and $j$~\cite{WeightShift}; additionally, $W_{ij} = 0$ for $i=j$. This yields a network that is complete (except for self-edges), weighted, and undirected. As for the interbank credit exposure network in Sec.~\ref{sec:interbank_network}, we use the edge weights for calculating CS values and their reciprocals for calculating PS and BC values.

Because an ETF is designed as a safe ``virtual stock'' that is a 
combination of individual stocks,
we expect ETFs to be correlated significantly with each other because they follow the market at large without as many wild fluctuations as individual stocks might exhibit. Naturally, they should also be correlated with their own constituents. We thus expect to observe a clear separation between core (ETF) and peripheral (individual stock) nodes.
As expected, the core nodes based on both CS and geodesic PS values are occupied by ETFs (see Table~\ref{SP500_ETFs_Top20}), although the correlation between CS and PS values is not very strong (see Table~\ref{SummaryTable_social1}). Note that even the weighted version
of geodesic BC is exactly the same for all of the nodes (and edges), so it is impossible to classify nodes or edges based on BC values.
This occurs because the (strict) triangle inequality $D_{ij} < D_{ik} + D_{kj}$ is satisfied for every triplet of nodes ($i,j,k$) in this fully connected network, so no indirect path ($i \to k \to j$) can ever be shorter than a direct path ($i \to j$).  
Therefore, this system illustrates that although BC values tend not to be very illuminating when a complete or almost complete network's edge weights are rather homogeneous, measuring node and edge coreness can still make it possible to quantify the importances of nodes and edges.


\subsection{United States Migration Network}
\label{sec:US_migration_network}

\begin{figure*}
\begin{tabular}{ll}
(a) & (b) \\
\hspace{.2in}
\includegraphics[width=0.90\columnwidth]{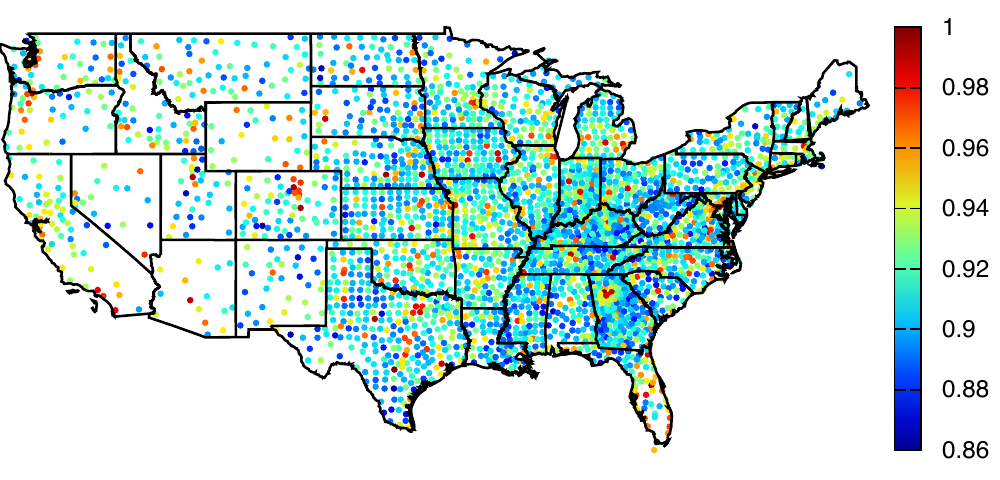} & 
\hspace{.2in}
\includegraphics[width=0.90\columnwidth]{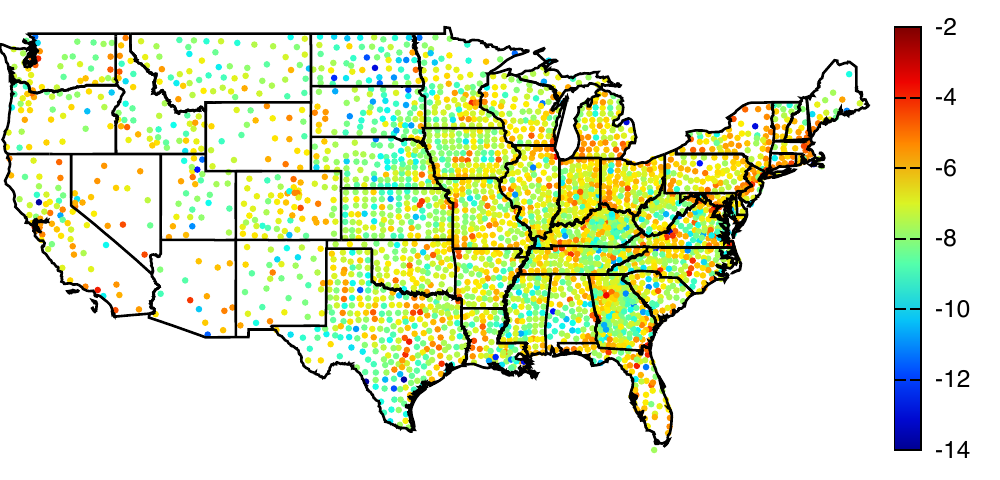} \\
(c) & (d) \\
\hspace{.2in}
\includegraphics[width=0.90\columnwidth]{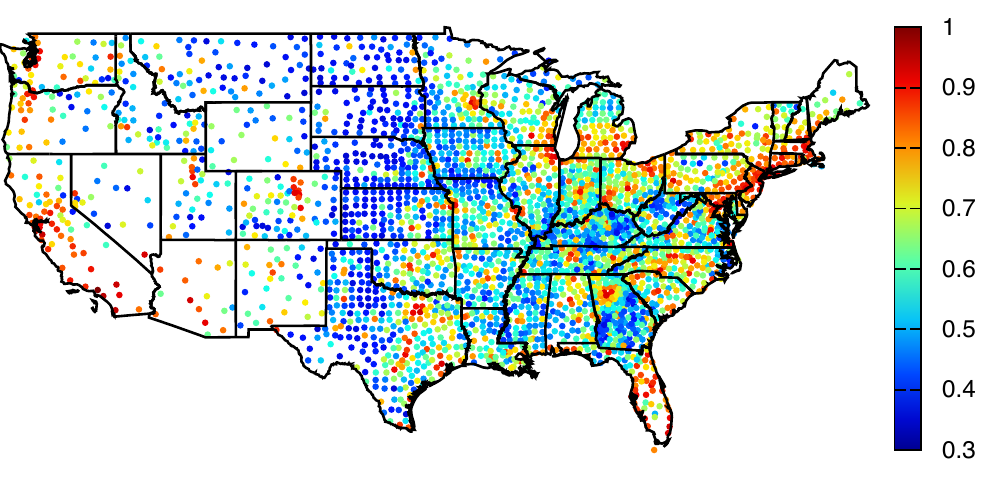} &
\hspace{.2in}
\includegraphics[width=0.90\columnwidth]{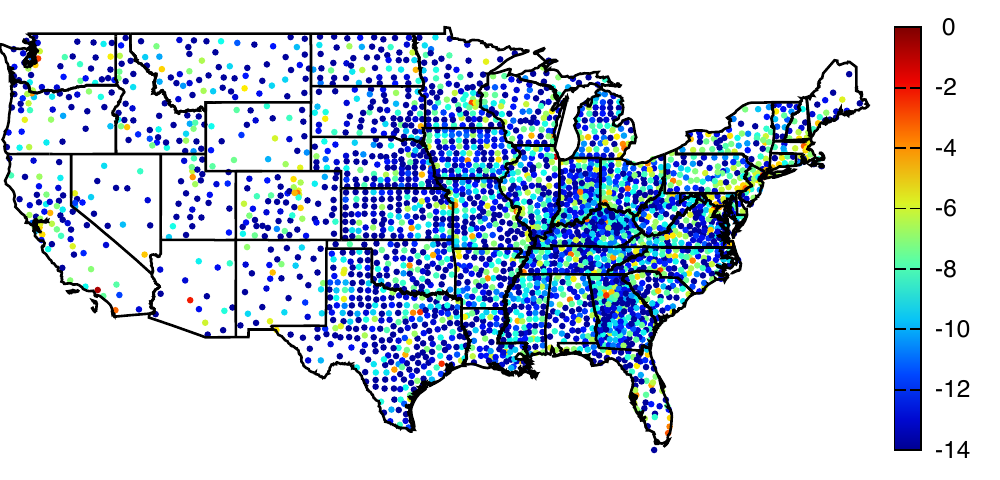}
\end{tabular}
\caption{Core values for (a,b) normalized flow ${\bf W}$ and (c,d) raw flow ${\bf W^{\textrm{raw}}}$ for migration between U.S. counties.  We color the counties according to their (a,c) CS values and (b,d) $\ln(\textrm{PS} + 10^{-6})$ values. We use the logarithm because of the heterogeneity in the PS values. We also indicate the state boundaries.
}
\label{US_migration_with_CS_PS}
\end{figure*}

\begin{table*}[t]
\caption{Top ten U.S. states with highest mean core scores (averaged over their component counties) for each combination of flow and coreness measure.  We calculate CS values and geodesic PS values for each county using both the normalized flow network ${\bf W}$ and raw flow network ${\bf W^{\textrm{raw}}}$; this yields four rank orderings. (We rounded all values to three significant digits.)
}
\label{US_migration_top10_table}
\begin{ruledtabular}
\begin{tabular}{rrrrr}
 & ${\bf W}$ & ${\bf W}$ & ${\bf W^{\textrm{raw}}}$ & ${\bf W^{\textrm{raw}}}$ \\
Rank & CS (value) & PS (value) & CS (value) & PS (value) \\
\hline
1 & Washington D.C. $(0.498)$ & Washington D.C. $(1.77 \times 10^{-1})$ & Washington D.C. $(0.883)$ & Washington D.C. $(3.22 \times 10^{-2})$ \\
2 & Arizona $(0.466)$ & Delaware $(5.30 \times 10^{-2})$ & New Jersey $(0.854)$ & California $(1.00 \times 10^{-2})$ \\
3 & New Jersey $(0.466)$ & Rhode Island $(3.10 \times 10^{-2})$ & Connecticut $(0.852)$ & Arizona $(7.90 \times 10^{-3})$ \\
4 & Florida $(0.466)$ & Connecticut $(2.09 \times 10^{-2})$ & Delaware $(0.825)$ & Massachusetts $(4.56 \times 10^{-3})$ \\
5 & Connecticut $(0.465)$ & New Hampshire $(1.64 \times 10^{-2})$ & Massachusetts $(0.809)$ & New York $(3.72 \times 10^{-3})$ \\
6 & California $(0.465)$ & Massachusetts $(1.23 \times 10^{-2})$ & Rhode Island $(0.794)$ & Illinois $(3.46 \times 10^{-3})$ \\
7 & Wyoming $(0.465)$ & Arizona $(1.13 \times 10^{-2})$ & California $(0.789)$ & Connecticut $(2.76 \times 10^{-3})$ \\
8 & Delaware $(0.463)$ & Vermont $(1.13 \times 10^{-2})$ & Arizona $(0.765)$ & Delaware $(2.68 \times 10^{-3})$ \\
9 & Oregon $(0.463)$ & Nevada $(1.02 \times 10^{-2})$ & New York $(0.764)$ & Florida $(2.33 \times 10^{-3})$ \\
10 & Maryland $(0.462)$ & Maine $(1.00 \times 10^{-2})$ & New Hampshire $(0.762)$ & Washington $(2.21 \times 10^{-3})$ \\
\end{tabular}
\end{ruledtabular}
\end{table*}

We now consider the United States (US) migration network between $3\,075$ counties in the mainland (i.e., excluding Alaska and Hawaii) during 1995--2000~\cite{Cucuringu2013,MigrationCensus,Perry2003}. 
We construct weighted, directed adjacency matrices using two types of flow measures: the raw values $W_{ij}^{\textrm{raw}}$ that represent the population that migrated from county $i$ to county $j$ and the normalized values $W_{ij} = {W_{ij}^{\textrm{raw}}} / \sqrt{P_i P_j}$ for the directed flow between the two counties,  
where $P_i$ is the total population of county $i$. In Fig.~\ref{US_migration_with_CS_PS}, we show the CS and geodesic PS values (with optimal paths that minimize the sum of the reciprocals of the weights) of the counties on a map of the U.S. The five counties with the largest CS values for the normalized adjacency matrix ${\bf W}$ are (in order) Bexar in Texas, Cobb in Georgia, Orange in Florida, Buffalo in Nebraska, and Boulder in Colorado. The five counties with the largest CS values for ${\bf W^{\textrm{raw}}}$ 
are (in order) Los Angeles in California, Orange in California, San Diego in California, Santa Clara in California, and Dallas in Texas.  There is a clear difference between our results for raw flow and normalized flow.

The five counties with the largest geodesic PS values for ${\bf W}$ are (in order) New York in New York, Chesapeake in Virginia, Washington D.C.,
Arlington in Virginia, and Fulton in Georgia.
The five counties with the largest geodesic PS values for ${\bf W^{\textrm{raw}}}$ are (in order) Los Angeles in California, Cook in Illinois, New York in New York,
Maricopa in Arizona, and Harris in Texas. We highlight the effect of county populations by comparing them with the CS and geodesic PS values. 
As shown in Table~\ref{SummaryTable_social2},
the correlation between CS and PS is much stronger for ${\bf W}$ than that for ${\bf W^{\textrm{raw}}}$,
so our two coreness values are more consistent more consistent with each other for the normalized flow than for the raw flow.
We also observe a correlation between coreness and county population for both ${\bf W}$ and ${\bf W^{\textrm{raw}}}$. (The correlation values are larger for the latter;
this is understandable, given the normalization by populations for the former.)
Therefore, even after the normalization of the flow by the populations of source and target counties, more populous counties also tend to be core counties (see Table~\ref{SummaryTable_social2}). As shown in Table~\ref{US_migration_top10_table}, the different choices of flow and coreness measures yield rather different results when aggregated at the state level (although Washington D.C. has the top coreness value in every case).  

It is useful to compare our observations to the intrastate versus interstate migration patterns that were discussed in Ref.~\cite{Cucuringu2013}, which reported that the top 14 states with maximum ``ratio degree'' (i.e., the ratio of incoming flux to outgoing flux) are (in order) Virginia, Michigan, Georgia, Indiana, Texas, Maine, New York, Missouri, Colorado, Louisiana, Mississippi, California, Ohio, and Wisconsin. Again, as discussed in Sec.~\ref{sec:structural_CP}, transportation-based coreness measures can help to characterize the importance of directed flow (which is population flow in this case).  Future work on directed versions of density-based coreness will be necessary to use those methods to help characterize core-source states versus core-sink states.


\section{Transportation networks}
\label{sec:examples_transporation_networks}

One expects many transportation networks to include core-periphery structures~\cite{Rombach2012}.  For example, metropolitan systems include both core and peripheral stations~\cite{marc-subway} and airline flight networks include high-traffic (i.e., hub) and low-traffic airports~\cite{dacosta}.  In this section, we examine core-periphery structure in several transportation networks.


\subsection{Rabbit Warren as a Three-Dimensional Road Network}
\label{sec:rabbit_warren_network}

\begin{table*}[t]
\caption{Pearson and Spearman correlation values between various core and centrality values (CS, PS, and BC) for transport and synthetic roadlike networks.  For the rabbit warren, we give (in parentheses) two-tailed $p$-values for the null hypothesis of absence of correlation. The values that we show for road networks (``Roads'') are the mean correlation values for all 100 roads, and we give standard errors in parentheses.
The results of 2D and 3D null-model, 100-node roadlike networks
are from an ensemble of 100 initial node locations. (In each case, we report mean values and standard errors over an ensemble.)
We use the same \textsc{SciPy} package in \textsc{Python}~\cite{SciPy} as in Tables~\ref{SummaryTable_social1} and~\ref{SummaryTable_social2}. 
}
\label{SummaryTable_transport}
\begin{ruledtabular}
\begin{tabular}{lllllllll}
Network & Correlation & CS vs PS & CS vs BC & PS vs BC & PS vs BC & CS vs GSNP & PS vs GSNP & PS vs GSNP \\
 & & (nodes) & (nodes) & (nodes) & (edges) & (nodes) & (nodes) & (edges) \\
\hline
Rabbit Warren~\cite{rabbit_warren_excavation_details} & Pearson & $0.231$ & $0.284$ & $0.561$ & $0.303$ & $0.371$ & $0.348$ & $8.25 \times 10^{-2}$ \\
 & & $(1.61 \times 10^{-2})$ & $(2.87 \times 10^{-3})$ & $(2.70 \times 10^{-10})$ & $(1.00 \times 10^{-3})$ & $(7.85 \times 10^{-5})$ & $(2.22 \times 10^{-4})$ & $(0.381)$ \\
\cline{2-9}
 & Spearman & $0.318$ & $0.437$ & $0.568$ & $0.403$ & $0.331$ & $0.293$ & $0.198$ \\
 & & $(8.08 \times 10^{-4})$ & $(2.29 \times 10^{-6})$ & $(1.48 \times 10^{-10})$ & $(7.84 \times 10^{-6})$ & $(4.69 \times 10^{-4})$ & $(2.09 \times 10^{-3})$ & $(3.42 \times 10^{-2})$ \\
\hline
3D Null Model~\cite{SHLee2013} & Pearson & $0.570(9)$ & $0.609(8)$ & $0.869(4)$ & $0.472(9)$ & $0.52(1)$ & $0.842(4)$ & $0.13(1)$ \\
\cline{2-9}
 & Spearman & $0.572(9)$ & $0.668(7)$ & $0.762(4)$ & $0.394(7)$ & $0.51(1)$ & $0.710(5)$ & $0.14(1)$ \\
\hline
Roads~\cite{SHLee2012} & Pearson & $-7(8) \times 10^{-4}$ & $-2(7) \times 10^{-4}$ & $4.1(4) \times 10^{-2}$ & $5.8(6) \times 10^{-2}$ & $-1(9) \times 10^{-4}$ & $4.2(4) \times 10^{-2}$ & $1.5(3) \times 10^{-2}$ \\
\cline{2-9}
 & Spearman & $-3(8) \times 10^{-4}$ & $4(7) \times 10^{-4}$ & $6.7(6) \times 10^{-2}$ & $8.8(8) \times 10^{-2}$ & $2(8) \times 10^{-4}$ & $6.0(6) \times 10^{-2}$ & $2.7(3) \times 10^{-2}$ \\
\hline
2D Null Model~\cite{SHLee2013} & Pearson & $0.29(2)$ & $0.33(1)$ & $0.668(7)$ & $0.247(9)$ & $0.25(2)$ & $0.675(7)$ & $9.7(8) \times 10^{-2}$ \\
\cline{2-9} 
 & Spearman & $0.36(1)$ & $0.45(1)$ & $0.683(6)$ & $0.288(8)$ & $0.31(2)$ & $0.692(6)$ & $0.216(9)$ \\
\hline
2D Null Model with & Pearson & $0.27(2)$ & $0.36(1)$ & $0.694(6)$ & $0.35(1)$ & $0.23(1)$ & $0.713(6)$ & $0.170(1)$ \\
\cline{2-9} 
Edge Crossing~\cite{SHLee2013} & Spearman & $0.35(2)$ & $0.46(1)$ & $0.707(6)$ & $0.389(9)$ & $0.31(2)$ & $0.692(6)$ & $0.216(9)$ \\
\end{tabular}
\end{ruledtabular}
\end{table*}

\begin{figure*}
\begin{tabular}{lll}
(a) & (b) & (c) \\
\hspace{.2in}
\includegraphics[height=0.27\textwidth]{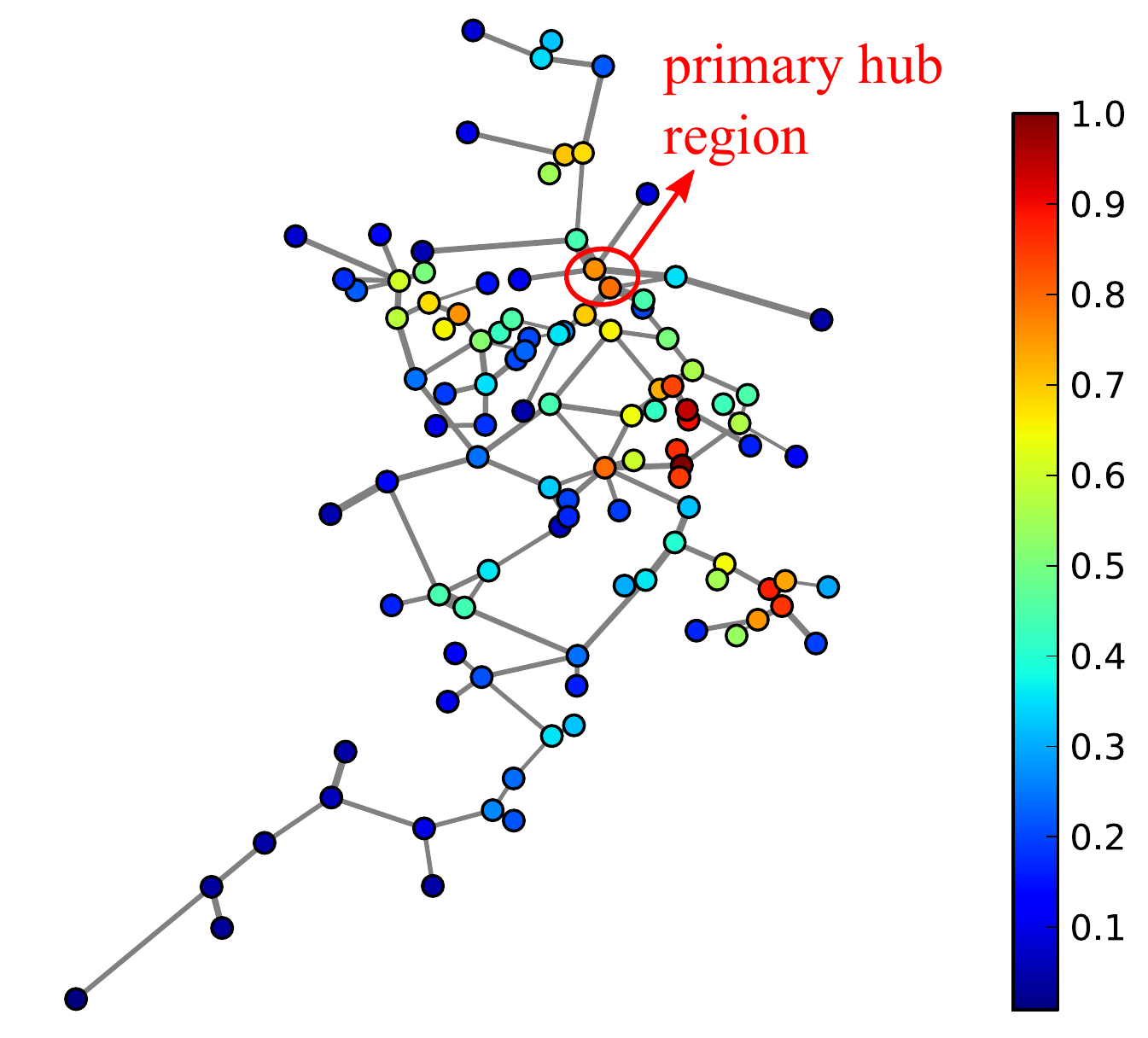} &
\hspace{.2in}
\includegraphics[height=0.27\textwidth]{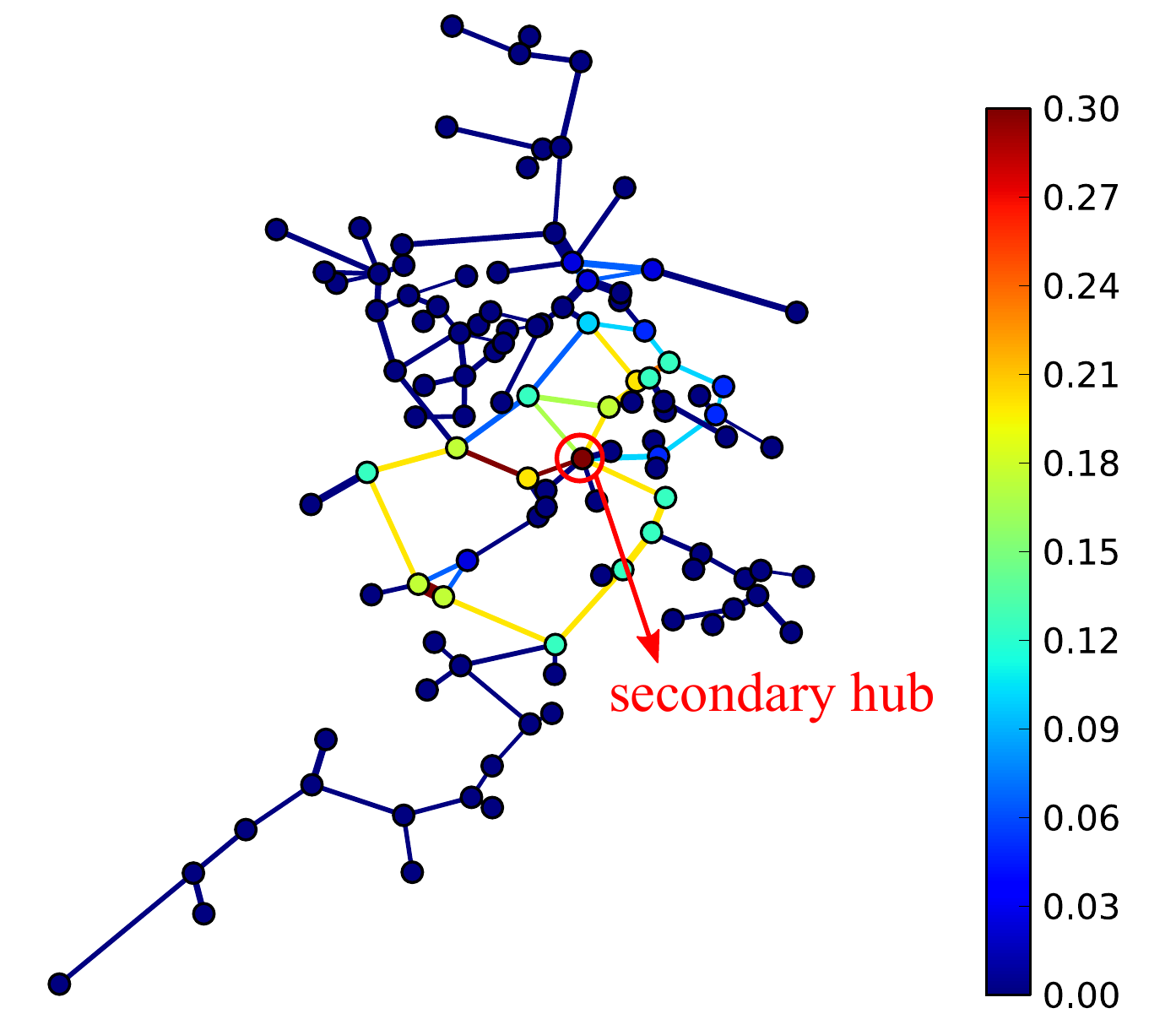} &
\hspace{.2in}
\includegraphics[height=0.27\textwidth]{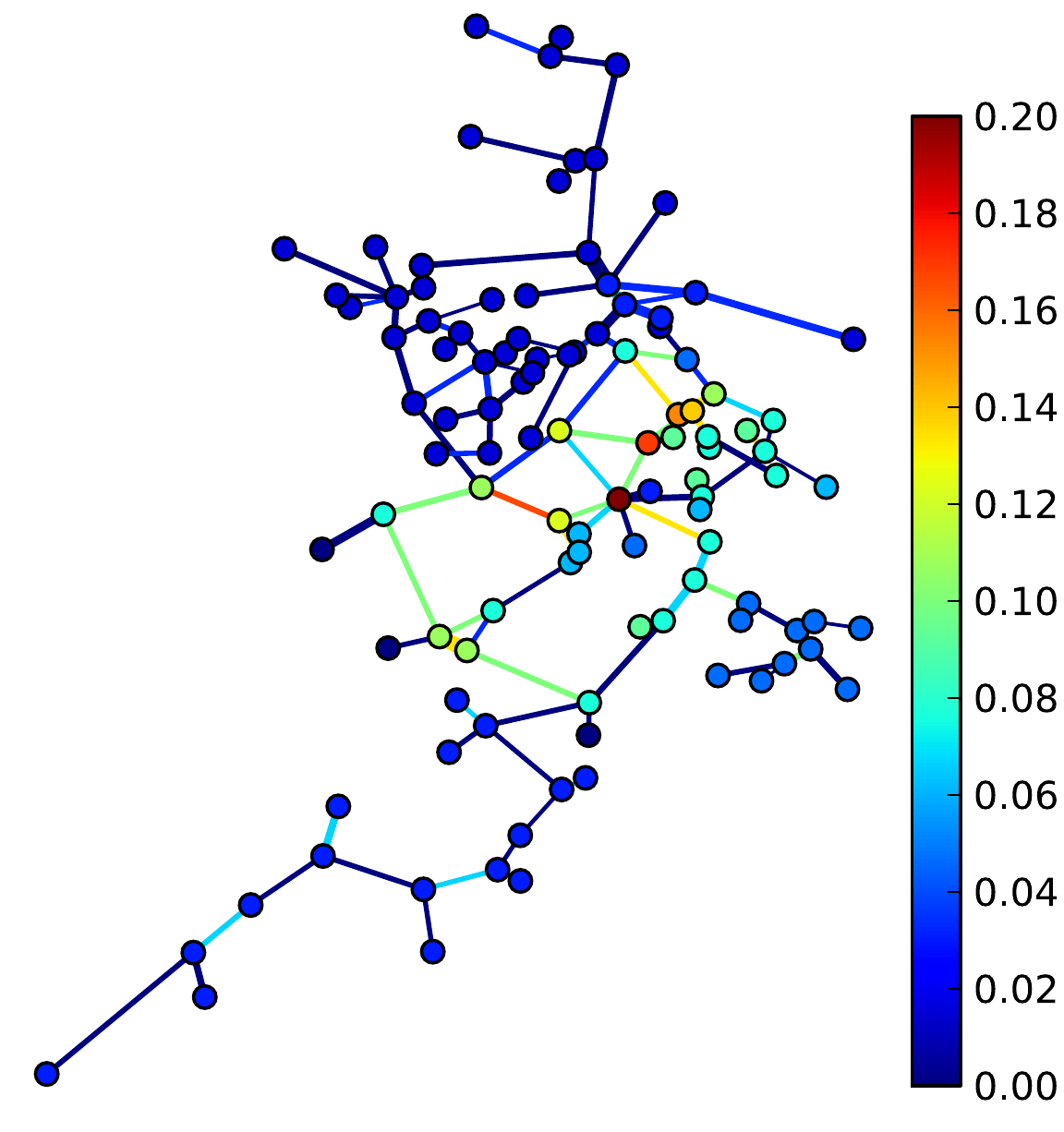} \\
\end{tabular}
\caption{Visualization of the rabbit-warren network. An edge's thickness is linearly proportional to the mean width of the tunnel segment that it represents. (It is difficult to discern the differences in width, as the widths are rather homogeneous.)
An edge's length is linearly proportional to its real length.
We color (a) the nodes according to CS values, (b) the nodes and edges according to geodesic PS values, and (c) the nodes and edges according to GSNP values.  We project the three-dimensional positions of nodes into a plane using a bird's-eye view.  The labels ``primary hub'' and ``secondary hub'' were applied by experts~\cite{BBC_documentary}, and the secondary hub was populated later in time than the primary one.  (The term ``primary'' is \emph{not} being used to indicate relative importance.)
}
\label{rabbit_warren_with_CS_PS}
\end{figure*}

The structure of animal burrows
is an important subject in zoology and animal behavior~\cite{Kolb1985,White2005}, and it is natural to view such structures through the lens of network science.  In this paper, we consider a European rabbit ({\it Oryctolagus cuniculus}) warren located in Bicton Gardens, Exeter, Devon, United Kingdom that was excavated~\cite{rabbit_warren_excavation_details} for the purpose of making a documentary series that was broadcast recently by the British Broadcasting Company (BBC)~\cite{BBC_documentary}.  We use a simplified network~\cite{RabbitWarrenData} that was generated from the detailed original three-dimensional (3D) warren structure in \textsc{ply} (Polygon File Format) by a researcher working with the BBC documentary team~\cite{SimonBuckley}.  

This network has 115 weighted, undirected edges that represent tunnel segments and 108 nodes that represent branching points or chambers, and we made this simplified network data public at~\cite{RabbitWarrenData}.  For the purpose of this paper, the weight of an edge is given by the reciprocal of the Euclidean distance between the two nodes that it connects, but one could use other information (such as the mean width of each individual tunnel segment) to define a set of weights. The 3D coordinates of the nodes are known, so this network gives a rare opportunity to investigate a transportation network that is used by animals.
The edge length is rather homogeneous (which is presumably deliberate), and the warren seems to have been 
developed in three phases via generational changes that are similar to an urban sprawl~\cite{DavidHosken}. 
In Fig.~\ref{rabbit_warren_with_CS_PS}, we show the rabbit-warren network projected into a two-dimensional (2D) plane.  In the figure panels, we color the nodes and edges according to various measures of coreness.

The node with the largest PS value in terms of both geodesic distance and GSN is the ``secondary hub'' marked in Fig.~\ref{rabbit_warren_with_CS_PS}(b) and was pointed out by an expert on rabbits.  The descriptor ``secondary'' refers to the fact that it was the second hub in temporal order; it is \emph{not} a statement of relative importance. The secondary hub has the second largest geodesic BC value. The ``primary hub'' region marked in Fig.~\ref{rabbit_warren_with_CS_PS}(a) has nodes with larger CS values than geodesic and GSNP values.  As one can see in Table~\ref{SummaryTable_transport}, the geodesic and GSNP values are highly correlated in the rabbit-warren network. According to the rabbit experts and the documentary~\cite{BBC_documentary}, stronger rabbits are able to acquire better breeding areas.  The best breeding areas experience lower traffic, and the breeding areas with the lowest PS values are the ones that the rabbit experts claimed are the best ones.   (If a breeding area experiences too much traffic, a rabbit needs to spend more time protecting its offspring to ensure that they are not killed by other rabbits~\cite{BBC_documentary}.)  Thus, coreness values seem to give insights about the structure of the rabbit warren that directly reflect aspects of the social hierarchy of rabbits. The breeding areas also have small BC values, so BC values are also insightful for the rabbit-warren network.

Additionally, as shown in Table~\ref{SummaryTable_transport}, the correlation between PS and BC 
values is much larger than that between CS and PS values and that between CS and BC values. This hints that PS values for a real transportation network are relevant for examining traffic in such a network. The PS and BC values of edges are also positively correlated.


\subsection{Urban Road Networks}
\label{sec:road_networks}

\begin{figure*}
\includegraphics[width=\textwidth]{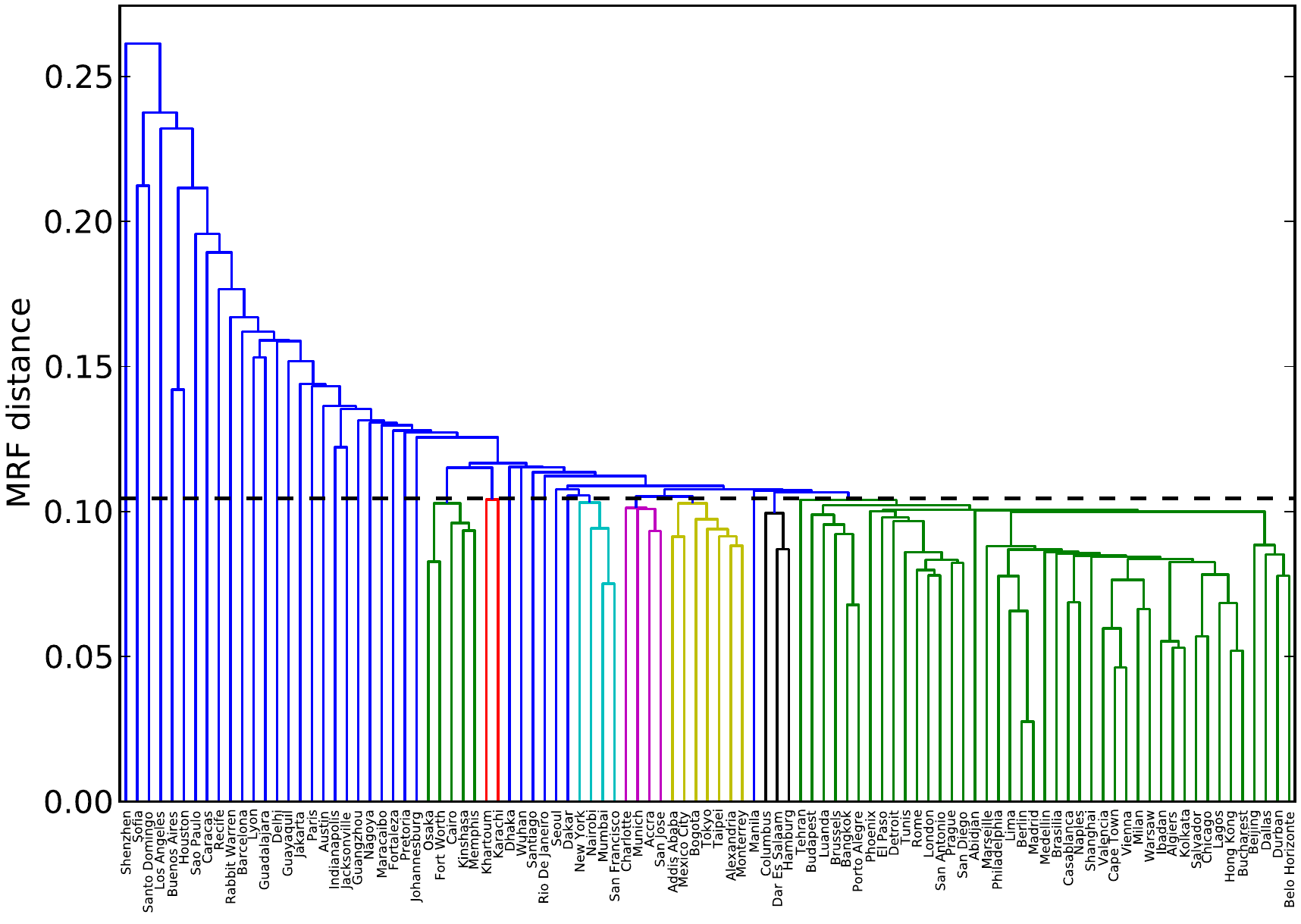} 
\caption{Taxonomy of the $100$ road networks and the rabbit warren using mesoscopic response functions (MRFs) based on community structure~\cite{Onnela2012}. The vertical axis gives a ``distance'' measured from the MRFs.  (This is analogous to a distance when studying phylogeny, so it indicates when different sets of networks diverge from each other in this taxonomy.) We set the threshold for assigning different colors to networks to be $40\%$ of the maximum distance (the dashed horizontal line) determined from the MRFs.
}
\label{taxonomy_road_networks}
\end{figure*}

To examine 2D transportation networks, we use road networks from $100$ large urban areas [square samples of the area (2 km $\times$ 2 km)] from all over the world~\cite{SHLee2012,RoadNetworkData}. To briefly compare these networks to each other (and to the rabbit warren, which is roadlike but embedded in 3D rather than 2D), we construct a taxonomy by using mesoscopic response functions (MRFs)~\cite{MRF_code} based on community structure~\cite{Onnela2012}.  As with the rabbit-warren network, the road networks are weighted and undirected, and the weight of each edge is given by the reciprocal of the Euclidean distance between the pair of nodes that it connects.  Thus, shorter roads correspond to stronger connections. We show the result of our taxonomy computation in Fig.~\ref{taxonomy_road_networks}. This taxonomy is based on pairwise closeness between networks determined from three types of normalized MRFs: a generalized modularity (i.e., with a resolution parameter)~\cite{CommunityReview} of network partitions, entropy of community sizes (based on their heterogeneity), and number of communities.  A network's MRF indicates how a particular quantity defined on a network partition changes as a function of a resolution parameter~\cite{Onnela2012}. As with the navigability measure in Ref.~\cite{SHLee2012}, the roads are not well-classified by external factors such as the continent in which cities are located. The rabbit warren is located between Recife and Barcelona in Fig.~\ref{taxonomy_road_networks}.

\begin{figure*}
\begin{tabular}{lll}
(a) & (b) & (c) \\
\hspace{.2in}
\includegraphics[width=0.28\textwidth]{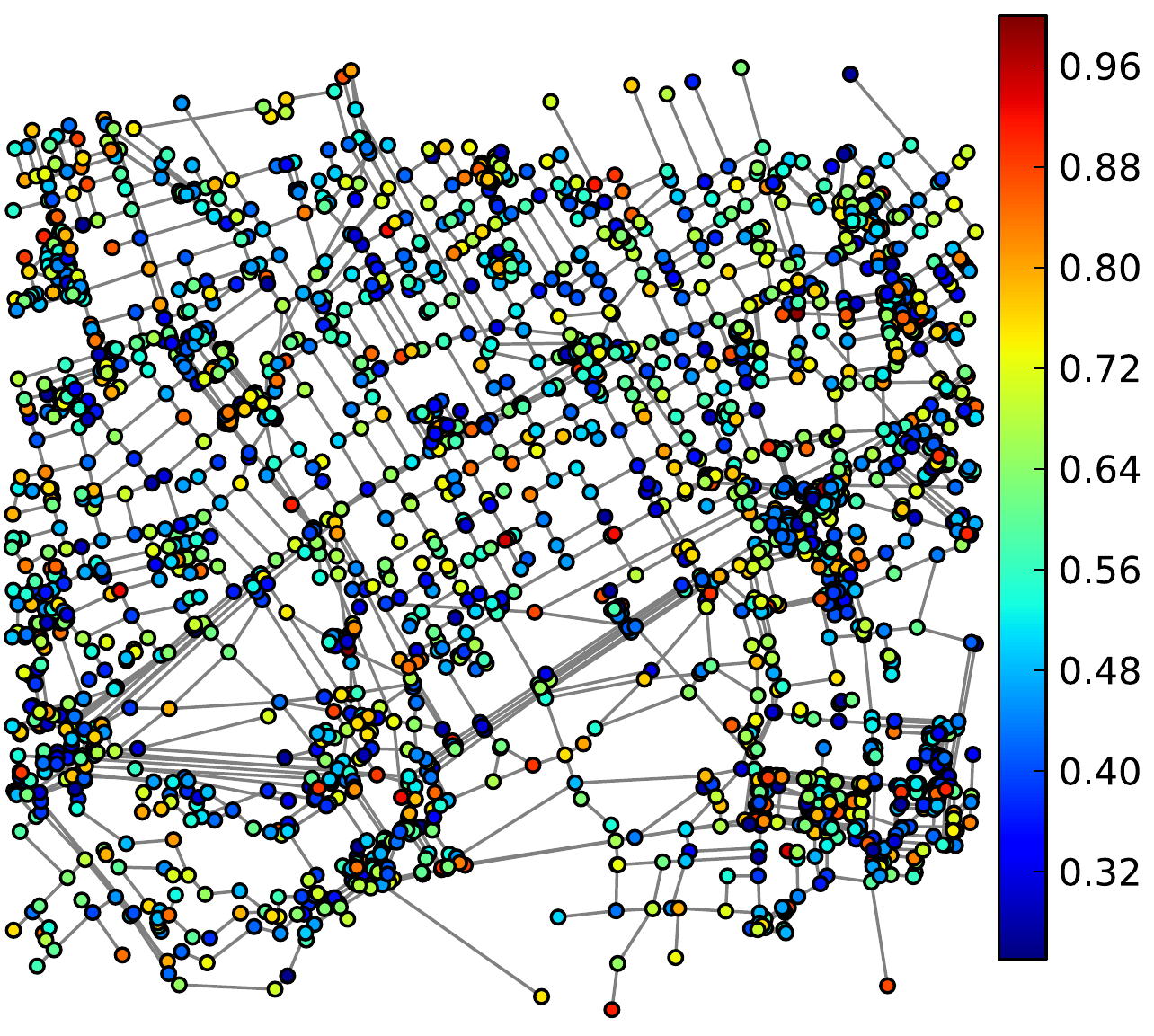} &
\hspace{.2in}
\includegraphics[width=0.28\textwidth]{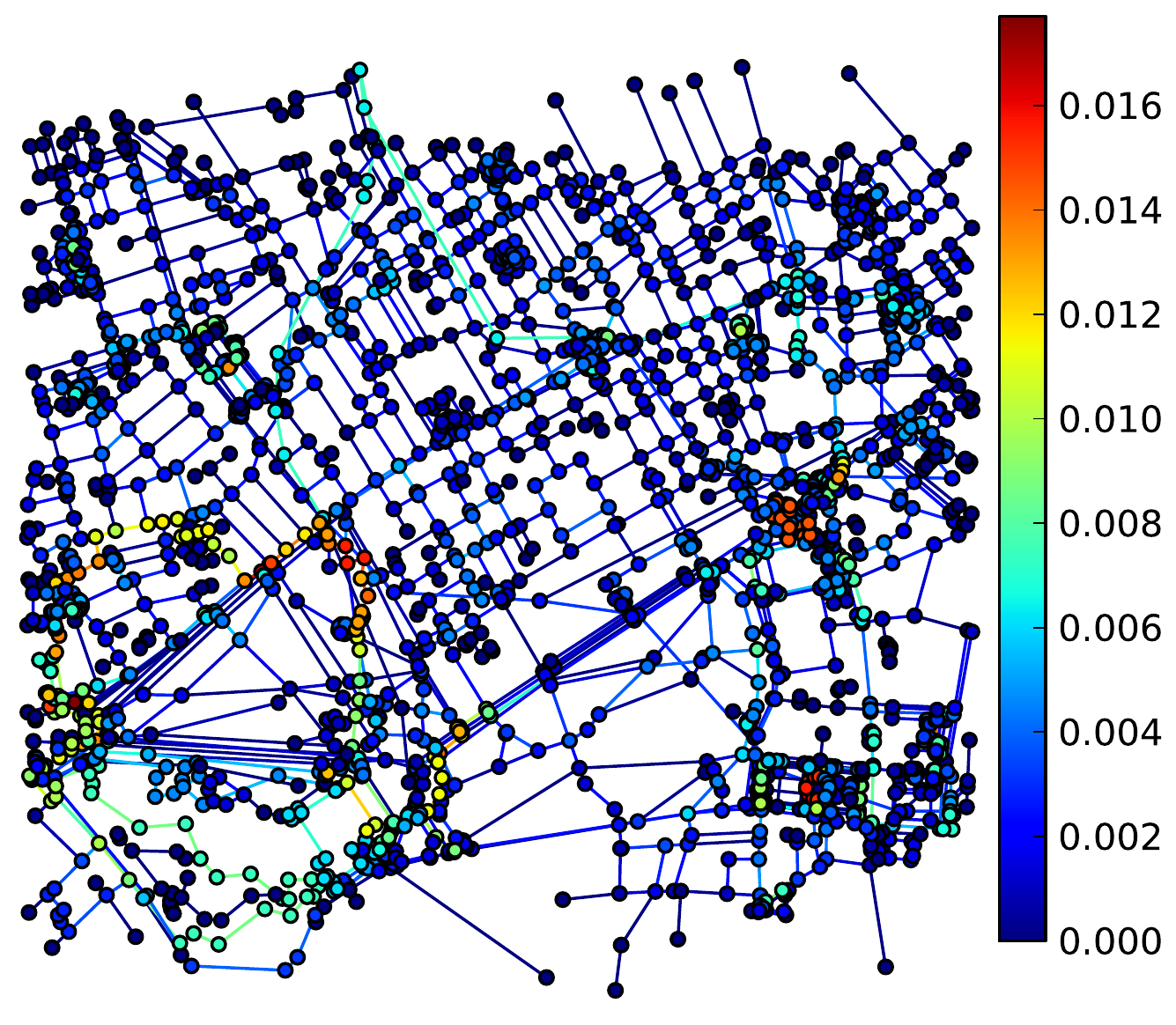} &
\hspace{.2in}
\includegraphics[width=0.28\textwidth]{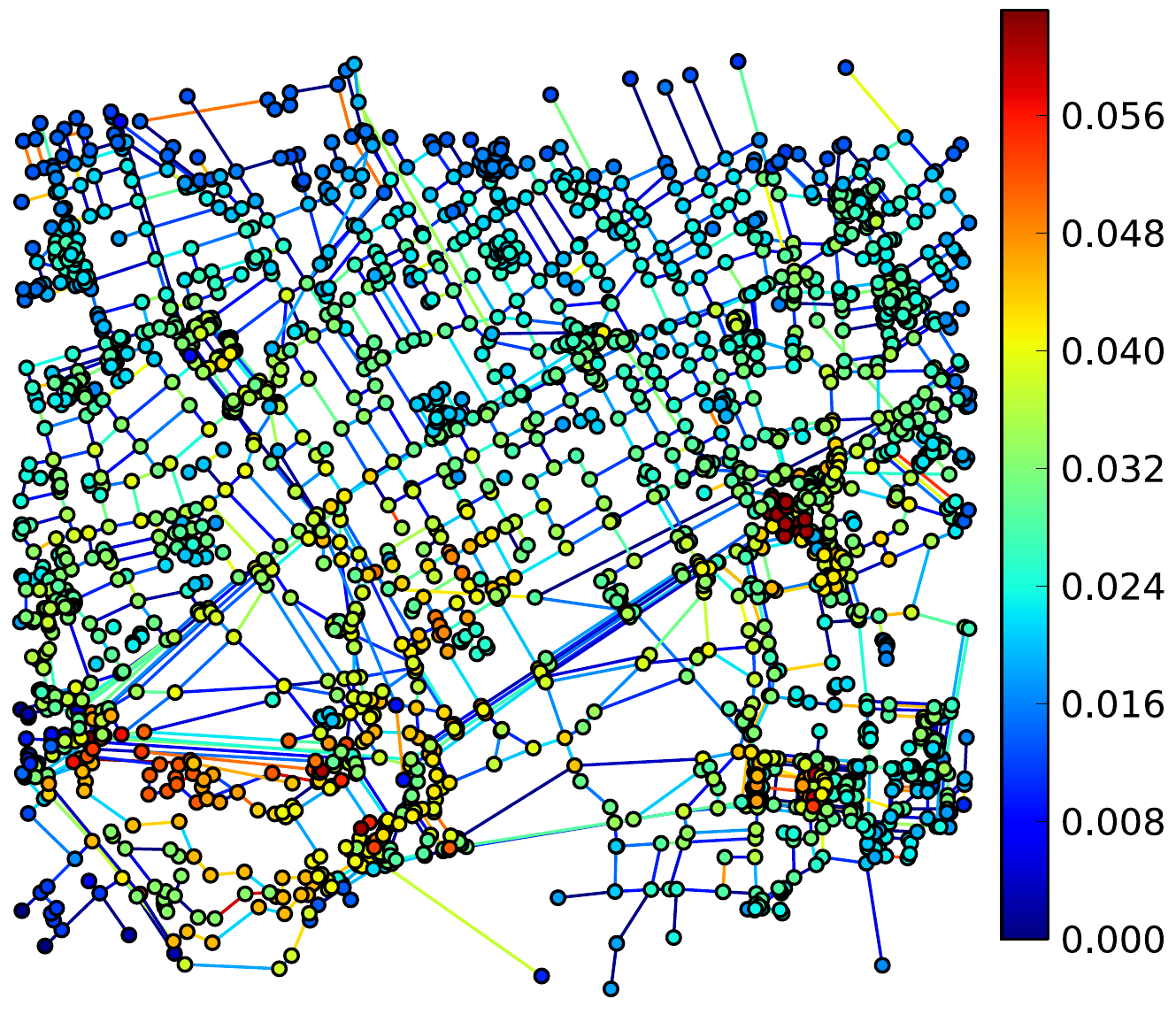} \\
\end{tabular}
\caption{A square sample (2 km $\times$ 2 km) of the road network in London. We color (a) the nodes based on their CS values, (b) the nodes and edges based on their geodesic PS values, and (c) the nodes and edges based on their GSNP values.
}
\label{London_with_CS_PS}
\end{figure*}

We examine core-periphery structure in the urban road networks. As an illustrative example, we show a square sample of the West End area of London in Fig.~\ref{London_with_CS_PS}. In Table~\ref{SummaryTable_transport}, we show correlation values between CS values, PS values, GSNP values, and BC values.
An interesting difference between the rabbit warren (which is embedded in 3D), which we discussed in Sec.~\ref{sec:rabbit_warren_network}, and the road networks (which are embedded in 2D) that one can see in this table is that the correlations of CS values versus other quantities (geodesic PS values, GSNP values, and BC values) are notably larger in the former.
It is natural to ask whether the smaller embedding dimension of the road networks as compared to the rabbit-warren network might be related to this property. The effects of spatial embeddedness on network structure is a difficult and interesting topic in general~\cite{marc-spatial}.  We thus investigate this possibility in more detail in Sec.~\ref{sec:null_model} by examining networks produced by generative models for 2D and 3D roadlike networks.


\subsection{Generative Models for 2D and 3D Road-Like Networks}
\label{sec:null_model}

\begin{figure*}
\begin{tabular}{lll}
(a) & (c) & (e) \\
\hspace{.1in}
\includegraphics[width=0.28\textwidth]{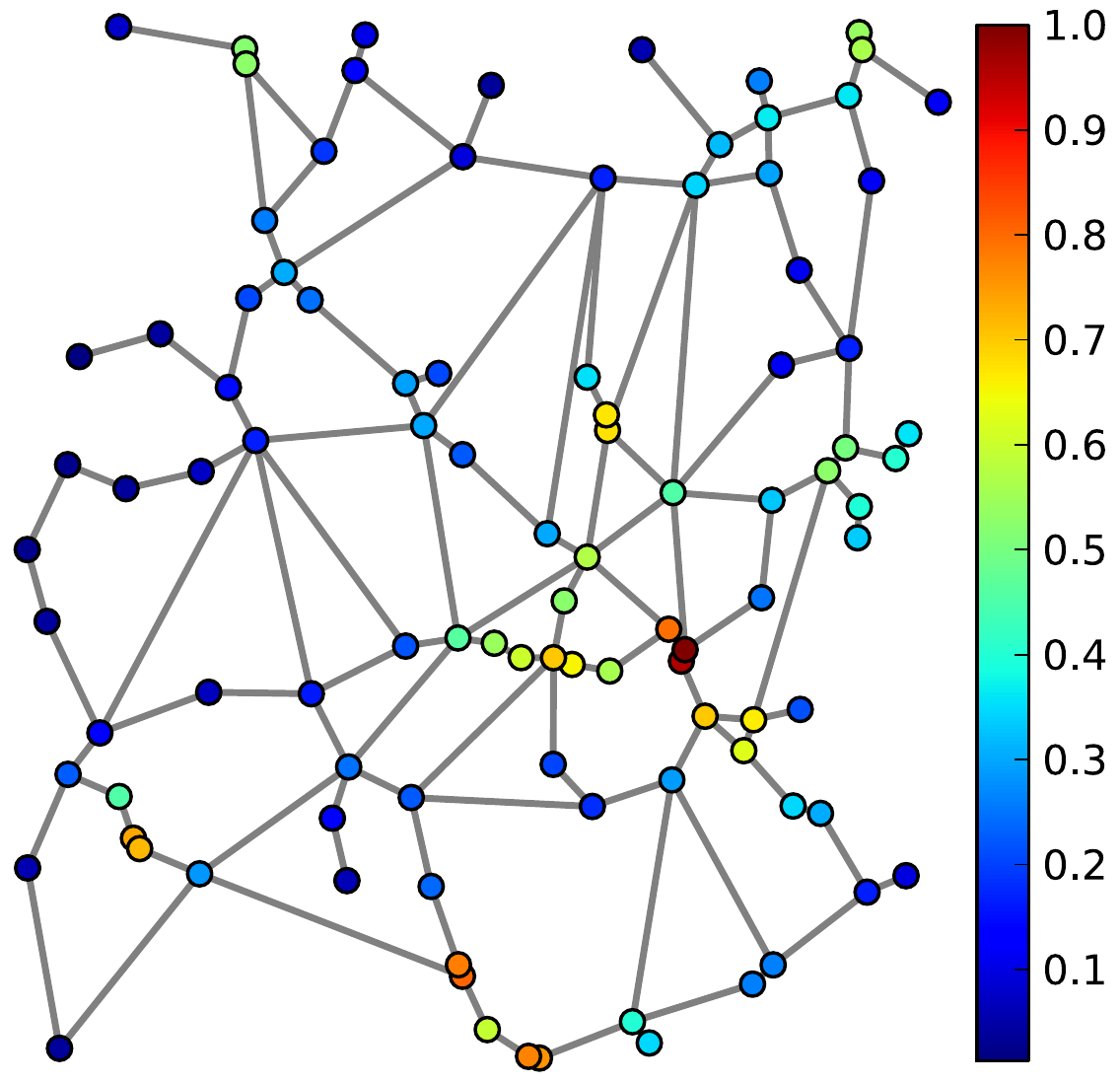} &
\hspace{.1in}
\includegraphics[width=0.28\textwidth]{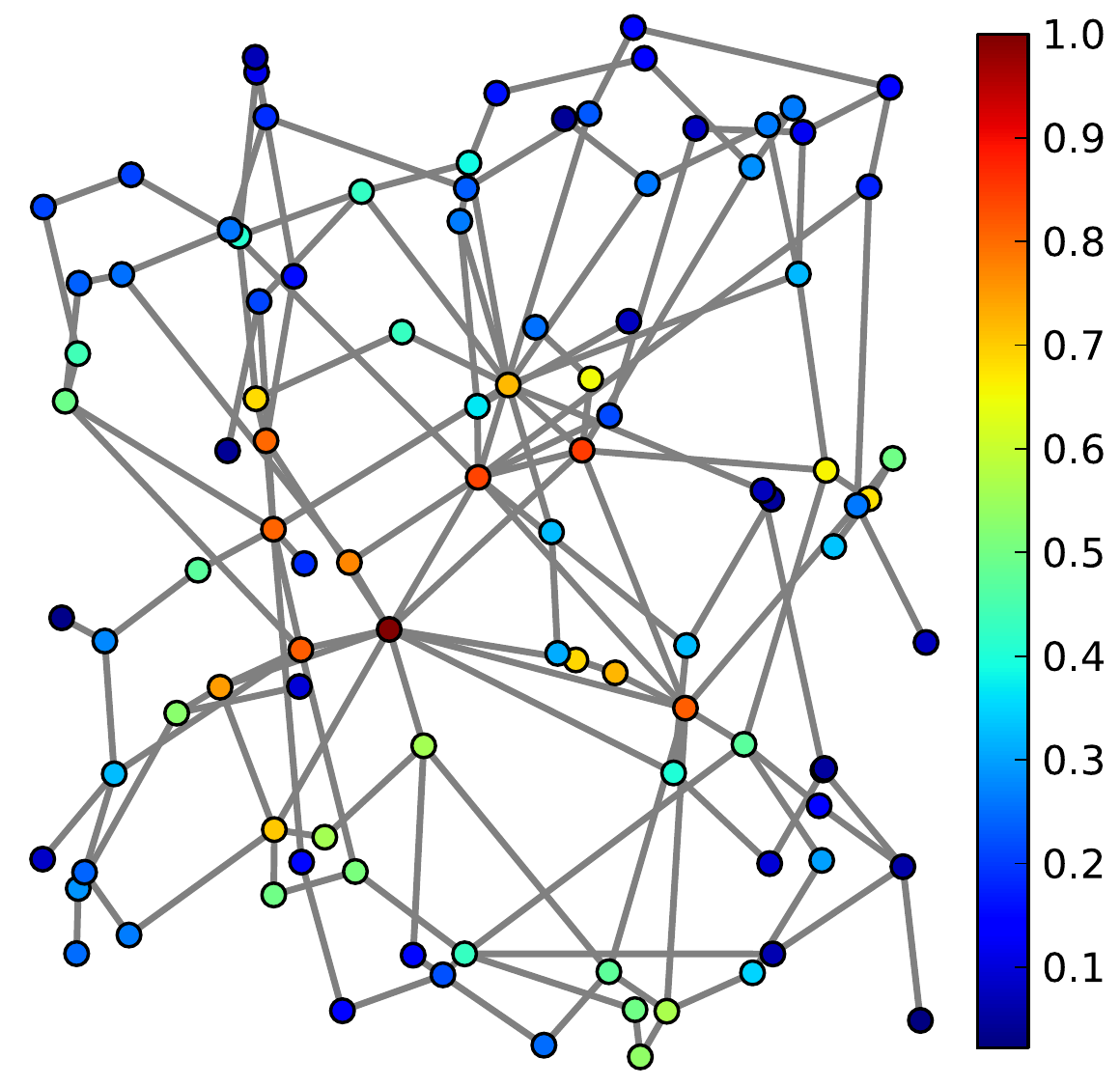} &
\hspace{.1in}
\includegraphics[width=0.28\textwidth]{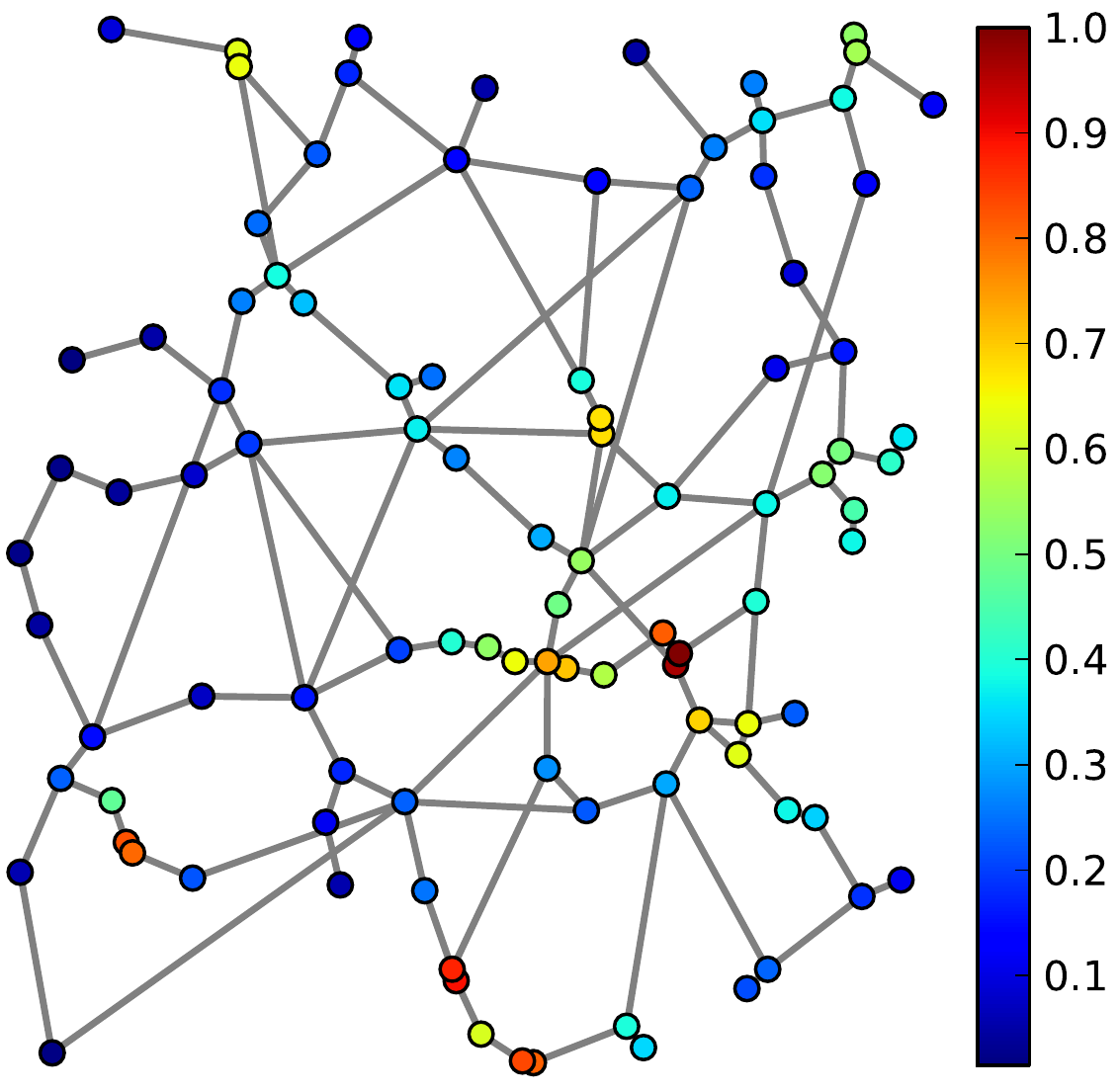} \\
(b) & (d) & (f) \\
\hspace{.1in}
\includegraphics[width=0.28\textwidth]{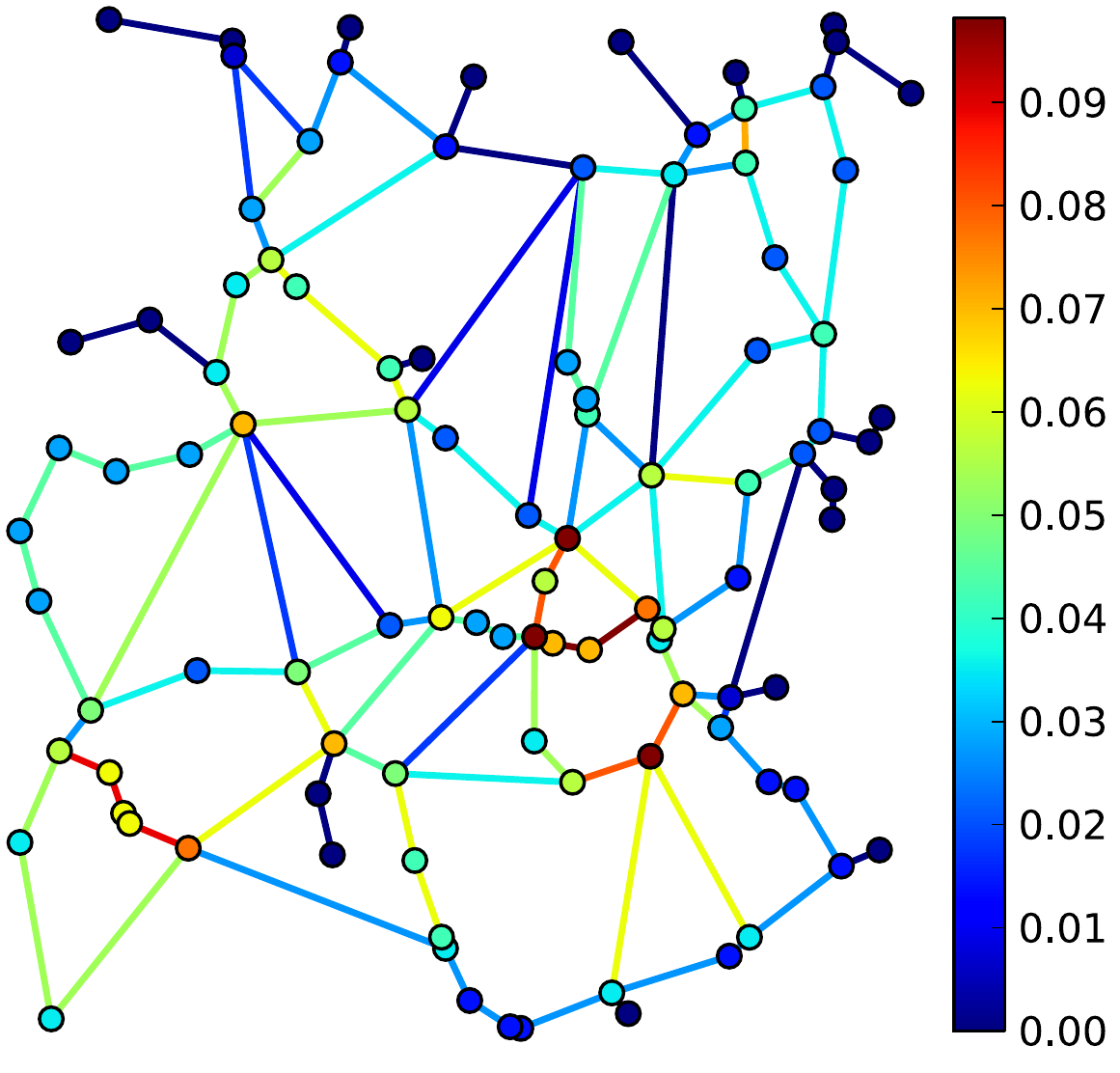} &
\hspace{.1in}
\includegraphics[width=0.28\textwidth]{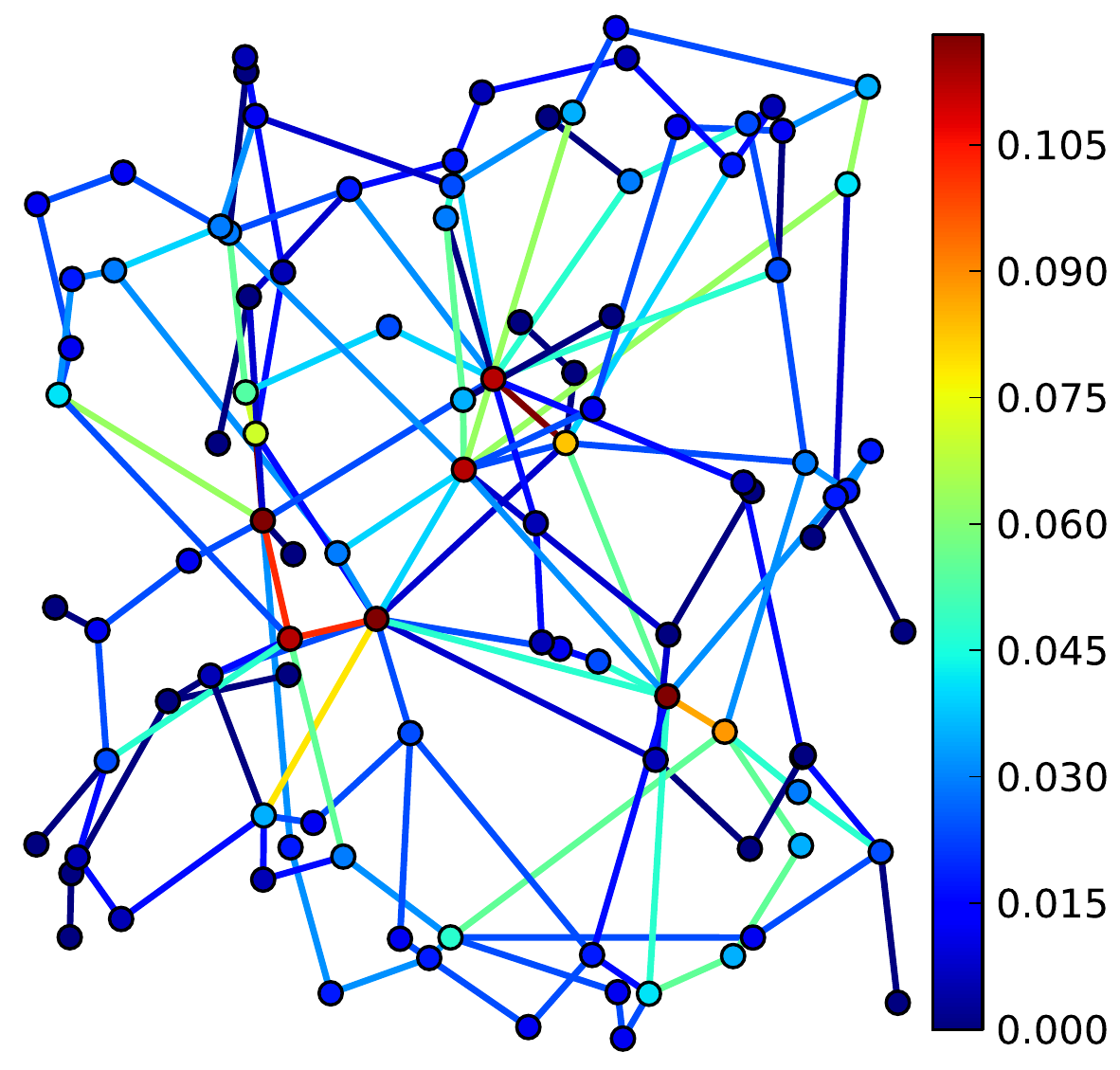} &
\hspace{.1in}
\includegraphics[width=0.28\textwidth]{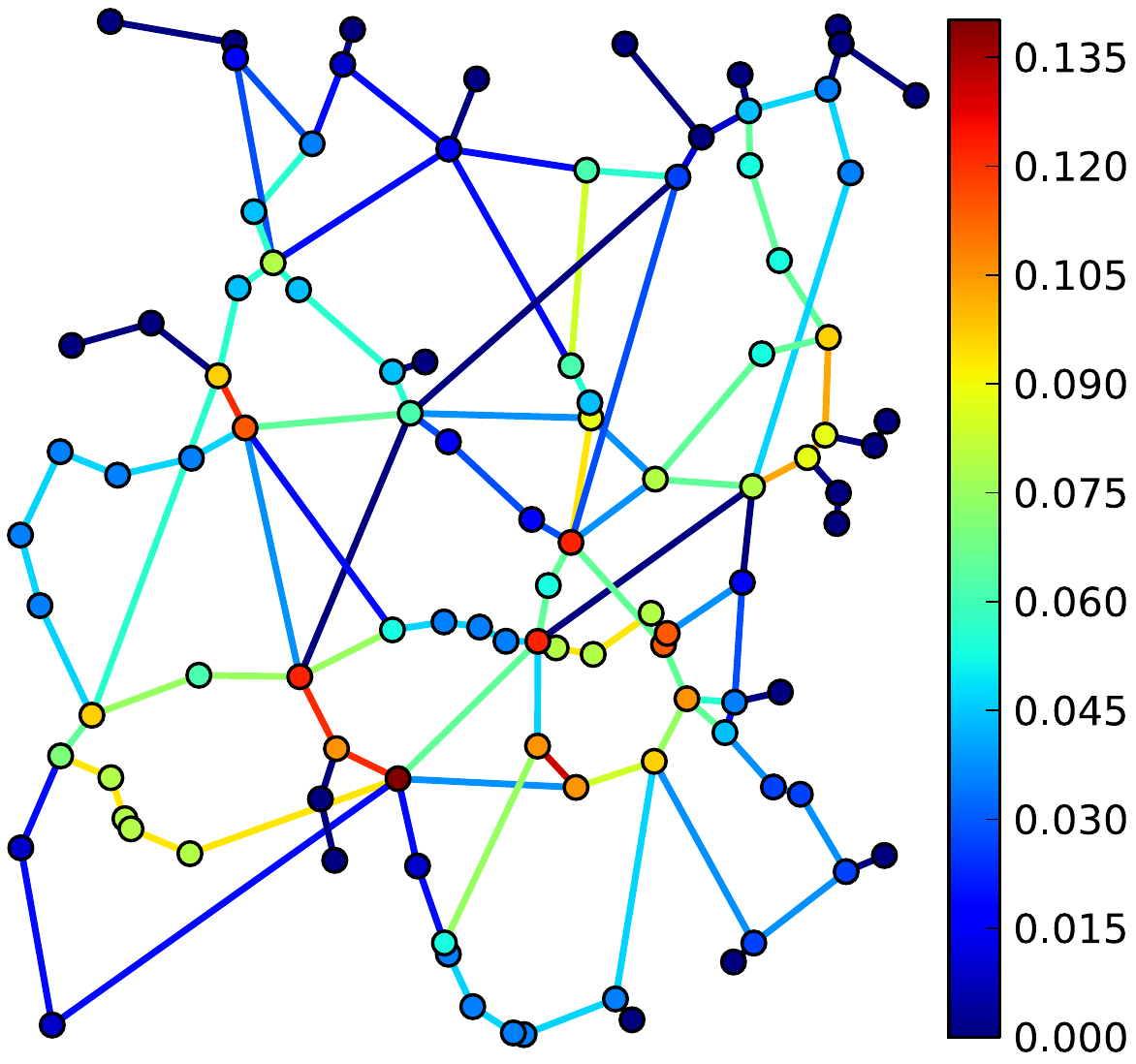} \\
\end{tabular}
\caption{Examples of 2D and 3D roadlike networks produced from generative models.  For the 2D example, we show (a) its nodes colored according to their CS values and (b) its nodes and edges colored according to their geodesic PS values.  We show the 3D example projected into a plane, and we color (c) its nodes by their CS values and (d) its nodes and edges by their PS values.  For these 3D networks, we add shortcuts greedily to make the mean path length as small as possible, and we note that the total length limit is twice the length of the minimum spanning tree (MST) determined from the initial node locations~\cite{SHLee2013}. The edges in panels (c) and (d) that appear to cross are, of course, artifacts of projecting the 3D network into a plane.  
For comparison, we also present (e) the CS values and (f) the PS values for a modified 2D null model in which crossed edges are allowed.  In this null model, we use the same initial locations of nodes as in panels (a) and (b). 
}
\label{null_CS_PS}
\end{figure*}

To examine correlations between the coreness measures and BC values in roadlike networks, we generate 2D and 3D roadlike structures from a recently introduced navigability-based model for road networks~\cite{SHLee2013}. We start by determining the locations of nodes either in the unit square (for 2D roadlike networks) or in the unit cube (for the 3D case).   We then add edges by constructing a minimum spanning tree (MST) via Kruskal's algorithm~\cite{KruskalsAlgorithm}.
Let $l_{\textrm{MST}}$ denote the total (Euclidean) length of the MST.
We then add the shortcut that minimizes the mean shortest path length over all node pairs, and we repeat this step until the total length of the network reaches a certain threshold. (When there is a tie, we pick one shortcut uniformly at random from the set of all shortcuts that minimize the shortest path length.)
Our final network is the set of nodes and edges right before the step that would force us to exceed this threshold by adding a new shortcut.  Reference~\cite{SHLee2013} called this procedure a ``greedy shortcut construction.'' In adding shortcuts, we also apply an additional constraint to emulate real road networks: new edges are not allowed to cross any existing edges. 

Consider a candidate edge $e_\textrm{cand}$ (among all of the possible pairs of nodes without an edge currently between them) that connects the vectors ${\bf q}$ and ${\bf q}+\Delta {\bf q}$.  We start by examining the 2D case.  Suppose that there is an edge $e_\textrm{ext}$ (which exists before the addition of a new shortcut) that connects ${\bf p}$ and ${\bf p} + \Delta {\bf p}$. The equation of intersection, 
\begin{equation*}
	{\bf p} + t \Delta {\bf p} = {\bf q} + u \Delta {\bf q}\,,
\end{equation*}	
then implies that
\begin{equation}
\begin{split}
	t &= \frac{[({\bf q} - {\bf p}) \times \Delta {\bf q} ]_z}{(\Delta {\bf p} \times \Delta {\bf q})_z}\,, \quad t \in [0,1]\,, \\
	u &= \frac{[({\bf q} - {\bf p}) \times \Delta {\bf p} ]_z}{(\Delta {\bf p} \times \Delta {\bf q})_z}\, \quad u \in [0,1]\,.
\end{split}
\label{intersection_solution}
\end{equation}
In Eq.~\eqref{intersection_solution}, the $z$ component (indicated by the subscripts) is perpendicular to the plane that contains the network.
If Eq.~\eqref{intersection_solution} has a solution, then $e_\textrm{ext}$
intersects with $e_\textrm{cand}$ , so $e_\textrm{cand}$ is excluded and we try another candidate edge. We continue until we exhaust every pair of nodes that are currently not connected to each other by an edge. We now consider the singular cases, in which the denominator in Eq.~\eqref{intersection_solution} equals $0$.
  When $\Delta {\bf p} \times \Delta {\bf q} = 0$, it follows that $\Delta {\bf p} \parallel \Delta {\bf q}$ (i.e., they are parallel to each other), so they cannot intersect; therefore, $e_\textrm{cand}$
is not excluded. When $\Delta {\bf p} \times \Delta {\bf q} = 0$ and $({\bf q} - {\bf p}) \times \Delta {\bf p} = 0$ [which is equivalent to $({\bf q} - {\bf p}) \times \Delta {\bf q} = 0$ because $\Delta {\bf p} \parallel \Delta {\bf q}$ implies that $({\bf q} - {\bf p})$, $\Delta {\bf p}$, and $\Delta {\bf q}$ are all parallel to each other], $e_\textrm{cand}$ and $e_\textrm{ext}$ are collinear and share infinitely many points, so
$e_\textrm{cand}$ is excluded from consideration in that case as well~\cite{GoldmanBook}.
We now consider the 3D case. The distance between (the closest points of) $e_\textrm{cand}$ and $e_\textrm{ext}$ is
\begin{equation*}
	d = \frac{\left|(\Delta {\bf p} \times \Delta {\bf q}) \cdot ({\bf q} - {\bf p})\right |} {\left|(\Delta {\bf p} \times \Delta {\bf q})\right|}\,.
\end{equation*}	
Thus, if $d > 0$, then it is guaranteed that $e_\textrm{cand}$ and $e_\textrm{ext}$ do not intersect. Again, $\Delta {\bf p} \times \Delta {\bf q} = 0$ corresponds to the parallel case, so $e_\textrm{cand}$ and $e_\textrm{ext}$ cannot intersect~\cite{GellertBook}. If $d = 0$, then the vectors ${\bf p}$ and ${\bf q}$ yield a plane,
so we obtain the same solution as in the 2D case, where we replace the $z$ component in Eq.~\eqref{intersection_solution} with the component that lies in the direction perpendicular to the relevant 2D plane.

We generate synthetic roadlike networks by placing 100 nodes uniformly at random inside of a unit square (2D) or cube (3D), and we use a threshold of $2 l_{\textrm{MST}}$ for the total length of the edges. In Fig.~\ref{null_CS_PS}, we show examples of 2D and 3D roadlike networks.  For each embedding dimension, we consider 50 different networks in our ensemble.  We consider 50 different initial node locations in each case, but that is the only source of stochasticity (except for another small source of stochasticity from the tie-breaking rule) because the construction process itself is deterministic. Our main observation from examining these synthetic networks is that correlations of CS values with other quantities (geodesic PS values, GSNP values, and BC values) are much larger in the 3D networks than in the 2D networks (see Table~\ref{SummaryTable_transport}).  This suggests that the embedding dimension of the roadlike networks is related to the correlations that we see in coreness (and betweenness) measures.

To further investigate the effects of the spatial embedding, we compare the results from the 2D generative model with a generative model that is the same except for a modified rule that allows some intersecting edges.  As shown in Table~\ref{SummaryTable_transport}, the correlation values between PS values and geodesic BC values for the modified model are slightly larger than in
the original model, though not that many edges cross each other in practice [see Figs.~\ref{null_CS_PS}(e) and (f)]. Therefore, although prohibiting edge crossings has some effect on correlations, the fact that most edges can be drawn in the same plane when edge crossings are allowed (i.e., the graphs in the modified model are ``almost 2D'' in some sense) suggests that the dimension in which a network (or most of a network) is embedded might have a larger effect on correlations between coreness (and betweenness) measures than the edge-crossing rule. 


\section{Conclusions and discussion}
\label{sec:discussion}

In this paper, we examined two types of core-periphery structure---one developed using intuition from social networks and another developed using intuition from transportation networks---in several networks from a diverse set of applications.  We showed that correlations between these different types of structures can be very different in different types of networks. This underscores the fact that it is important to develop different notions of core-periphery structure that are appropriate for different situations.  We also illustrated in our case studies that coreness measures can detect important nodes and edges.  For roadlike networks, we also examined the effect of spatial embeddedness on correlations between coreness measures.

As with the study of community structure (and many other network concepts), the notion of core-periphery structure is context-dependent.  For example, we illustrated that the intuition behind what one considers a core road or junction in a road (or roadlike) network is different from the intuition behind what one considers to be a core node in a social network.  Consequently, it is important to develop and investigate (and examine correlations between) different notions of core-periphery structure.  We have taken a step in this direction through our case studies in this paper, and we also obtained insights in several applications.  Our work also raises interesting questions.  For example, how much of the structure of the rabbit warren stems from the fact that it is embedded in 3D, how much of its structure stems from its roadlike nature, and how much of its structure depends fundamentally on the fact that it was created by rabbits (but would be different from other roadlike networks that are also embedded in 3D)?

Finally, we emphasize that core-periphery structure is a fascinating and important aspect of networks that deserves much more attention than it has received thus far in the literature.

\begin{acknowledgements}

S.H.L. and M.A.P. were supported by Grant No. EP/J001795/1 from the EPSRC, and M.A.P. was also supported by the European Commission FET-Proactive project PLEXMATH (Grant No. 317614). M.C. was supported by AFOSR MURI Grant No. FA9550-10-1-0569. Some computations were carried out in part using the servers and computing clusters in the Complex Systems and Statistical Physics Lab (CSSPL) at the Korea Advanced Institute of Science and Technology (KAIST). David Lusseau provided the dolphin network data, Hannah Sneyd and Owen Gower provided the rabbit-warren data, and Davide Cellai provided the interbank network data.  We thank Puck Rombach for the code to produce CS values and Dan Fenn for the code to produce MRFs.  We thank Simon Buckley and John Howell for assistance with preparation of the rabbit-warren data. We thank Young-Ho Eom and Taha Yasseri for helpful comments on the interbank network. Finally, we thank Roger Trout and Anne McBride for their expert opinions on the rabbit warren.

\end{acknowledgements}


\begin{thebibliography}{00}
\bibitem{ComplexNetwork}
S.\,N. Dorogovtsev and J.\,F.\,F. Mendes, Adv. Phys. {\bf 51}, 1079 (2002);
S. Boccaletti, V. Latora, Y. Moreno, M. Chavez, and D.-U. Hwang,
Phys. Rep. {\bf 424}, 175 (2006);
M. E. J. Newman, \emph{Networks: An Introduction} (Oxford University Press, Oxford, U.K., 2010);
S. Wasserman and K. Faust, \emph{Social Network Analysis: Methods and Applications} (Cambridge University Press, Cambridge, U.K., 1994).


\bibitem{CommunityReview} M.\,A. Porter, J.-P. Onnela, and P.\,J. Mucha, Not. Am. Math. Soc. {\bf 56}, 1082 (2009); S. Fortunato, Phys. Rep. {\bf 486}, 75 (2010).

\bibitem{ng2004} M. E. J. Newman and M. Girvan, Phys. Rev. E {\bf 69}, 026113 (2004).


\bibitem{YYAhn2010} Y.-Y. Ahn, J.\,P. Bagrow, and S. Lehmann, Nature (London) {\bf 466}, 761 (2010).

\bibitem{Rosvall2007} M. Rosvall and C.\,T. Bergstrom, Proc. Natl. Acad. Sci. USA {\bf 104}, 7327 (2007); {\it ibid}. {\bf 105}, 1118 (2008).

\bibitem{lambiotte08} R. Lambiotte, J.-C. Delvenne, and M. Barahona, arXiv:0812.1770.

\bibitem{jeub2013} L. G. S. Jeub, P. Balichandran, M. A. Porter, P. J. Mucha, and M. W. Mahoney, arXiv:1403.3795.

\bibitem{anna2010} A. C. F. Lewis, N. S. Jones, M. A. Porter, and C. M. Deane. BMC Sys. Bio. {\bf 4}, 100 (2010).

\bibitem{dani2011} D. S. Bassett, N. F. Wymbs, M. A. Porter, P. J. Mucha, J. M. Carlson, and S. T. Grafton, Proc. Natl. Acad. Sci. USA {\bf 108}, 7641 (2011).

\bibitem{mason2005} M. A. Porter, P. J. Mucha, M. E. J. Newman, and C. M. Warmbrand, Proc. Natl. Acad. Sci. USA {\bf 102}, 7057 (2005).


\bibitem{mucha2010} P. J. Mucha, T. Richardson, K. Macon, M. A. Porter, and J.-P. Onnela, Science {\bf 328}, 876 (2010).

\bibitem{marta2007} M. C. Gonz{\'a}lez, H. J. Herrmann, J. Kert{\'e}sz, and T. Vicsek, Physica A {\bf 379}, 307 (2007).

\bibitem{traud2012} A. L. Traud, P. J. Mucha, and M. A. Porter, Physica A {\bf 391}, 4165 (2012).

\bibitem{everettrole} M. G. Everett and S. P. Borgatti, J. Math. Sociol. {\bf 19}, 29 (1994).

\bibitem{doreian} P. Doreian, V. Batagelj, and A. Ferligoj, \emph{Generalized Blockmodeling} (Cambridge University Press, Cambridge, U.K., 2004).

\bibitem{Borgatti1999} S.\,P. Borgatti and M.\,G. Everett, Soc. Networks {\bf 21}, 375 (1999).

\bibitem{Holme2005} P. Holme, Phys. Rev. E {\bf 72}, 046111 (2005).

\bibitem{Silva2008} M.\,R. da Silva, H. Ma, and A.-P. Zeng, Proc. IEEE {\bf 96}, 1411 (2008).

\bibitem{Rombach2012} M.\,P. Rombach, M.\,A. Porter, J.\,H. Fowler, and P.\,J. Mucha, SIAM J. App. Math. {\bf 74}, 167 (2014).

\bibitem{Csermely2013} P. Csermely, A. London, L.-Y. Wu, and B. Uzzi, J. Complex Networks {\bf 1}, 93 (2013).

\bibitem{JYang2012} J. Yang and J. Leskovec, arXiv:1205.6228.

\bibitem{wallerstein} I. Wallerstein, \emph{The Modern World-System} (Academic, New York, 1974).

\bibitem{bascompte} J. Bascompte, P. Jordano, C. J. Meli\'an, and J. M. Olesen, Proc. Natl. Acad. Sci. USA {\bf 100}, 9383 (2003).

\bibitem{Cucuringu2013a} M. Cucuringu, M.\,P. Rombach, S.\,H. Lee, and M.\,A. Porter (unpublished).

\bibitem{Onnela2012} J.-P. Onnela, D.\,J. Fenn, S. Reid, M.\,A. Porter, P.\,J. Mucha, M.\,D. Ficker, and N.\,S. Jones, Phys. Rev. E {\bf 86}, 036104 (2012).

\bibitem{Newman2006} M. E. J. Newman, Phys. Rev. E {\bf 74}, 036104 (2006).

\bibitem{Mahoney2009} J. Leskovec, K. J. Lang, A. Dasgupta, and M. W. Mahoney, Internet Math. {\bf 6}, 29 (2009).

\bibitem{Bassett2013} D.\,S. Bassett, N.\,F. Wymbs, M.\,P. Rombach, M.\,A. Porter, P.\,J. Mucha, and S.\,T. Grafton,
PLOS Comput. Biol. {\bf 9}, e1003171 (2013).

\bibitem{Kirkpatrick1983} S. Kirkpatrick, C.\,D. Gelatt, Jr., and M.\,P. Vecchi, Science {\bf 220}, 671 (1983).

\bibitem{Freeman1977} L.\,C. Freeman, Sociometry {\bf 40}, 35 (1977).

\bibitem{KIGoh2001} K.-I. Goh, B. Kahng, and D. Kim, Phys. Rev. Lett. {\bf 87}, 278701 (2001).

\bibitem{rwcore} F. Della Rosso, F. Dercole, and C. Piccardi, Sci. Rep. {\bf 3}, 1467 (2013).

\bibitem{Girvan2002} M. Girvan and M.\,E.\,J. Newman, Proc. Natl. Acad. Sci. USA {\bf 99}, 7821 (2002).

\bibitem{marc2012} E. Strano, V. Nicosia, V. Latora, S. Porta, and M. Barthelemy, Sci. Rep. {\bf 2}, 296 (2012).

\bibitem{eduardo} E. L\'opez, R. Parshani, R. Cohen, S. Carmi, and S. Havlin, Phys. Rev. Lett. {\bf 99}, 188701 (2007).

\bibitem{rwbetween} M. E. J. Newman, Social Networks {\bf 27}, 39 (2005).

\bibitem{SHLee2012} S.\,H. Lee and P. Holme, Phys. Rev. Lett. {\bf 108}, 128701 (2012).

\bibitem{GSN_details} For details, see Ref.~\cite{SHLee2012}, which defines a GSN for $d = 2$. However, it is easy to generalize GSNs to $\mathbb{R}^d$ for any integer $d \geq 2$.

\bibitem{Lusseau2003} D. Lusseau, K. Schneider, O.\,J. Boisseau, P. Haase, E. Slooten, and S.\,M. Dawson,
Behav. Ecol. Sociobiol. {\bf 54}, 396 (2003); D. Lusseau, Proc. R. Soc. Lond. B {\bf 270}, S186 (2003).

\bibitem{NewmanWebsite} The network data is downloadable from Mark Newman's website: \url{http://www-personal.umich.edu/~mejn/netdata/}

\bibitem{Kamada1989} T. Kamada and S. Kawai, Inf. Process. Lett. {\bf 31}, 7 (1989).

\bibitem{graphviz} Copyright 2013, \textsc{NetworkX} Developers (last updated in 2013), \textsc{graphviz\_layout}: Create node positions using \textsc{Pydot} and \textsc{Graphviz}. \url{http://networkx.github.io/documentation/latest/reference/generated/networkx.drawing.nx_pydot.graphviz_layout.html}

\bibitem{Lusseau2004} D. Lusseau and M.\,E.\,J. Newman, Proc. R. Soc. Lond. B (Suppl.) {\bf 270}, S477 (2004).

\bibitem{SciPy} SciPy.org, \url{http://www.scipy.org/}

\bibitem{Lusseau2007} D. Lusseau, Evol. Ecol. {\bf 21}, 357 (2007).

\bibitem{Lusseau2006} D. Lusseau, Behav. Processes {\bf 73}, 257 (2006).

\bibitem{Haldane2011} A.\,G. Haldane and R.\,M. May, Nature (London) {\bf 469}, 351 (2011).

\bibitem{Lloyd2013} S. Lloyd, arXiv:1302.3199.

\bibitem{CraigWorkingPaper} B. Craig and G. von Peter, 
\url{http://www.bis.org/publ/work322.pdf} (2010).

\bibitem{LuxWorkingPaper} T. Lux and D. Fricke,
\url{http://ideas.repec.org/p/kie/kieliw/1759.html} (2012).

\bibitem{CreditExposure} Credit Exposure, \url{http://www.investopedia.com/terms/c/credit-exposure.asp}

\bibitem{EBA} European Banking Authority, \url{http://www.eba.europa.eu/}

\bibitem{InterbankNetwork} 2011 EU-wide stress test results. \url{http://eba.europa.eu/risk-analysis-and-data/eu-wide-stress-testing/2011/results}

\bibitem{FinancialStabilityReview} European Central Bank, {\it Financial Stability Review June 2012} (European Central Bank, Frankfurt am Main, 2012).
\url{http://www.ecb.europa.eu/pub/pdf/other/financialstabilityreview201206en.pdf}

\bibitem{GooglingSocialInteractions} S.\,H. Lee, P.-J. Kim, Y.-Y. Ahn, and H. Jeong,
PLOS ONE {\bf 5}, e11233 (2010).

\bibitem{NetworkXEdges} Copyright 2010, \textsc{NetworkX} Developers (late updated in 2012), \textsc{draw\_networkx\_edges}: Draw the edges of the graph $G$. \url{http://networkx.lanl.gov/reference/generated/networkx.drawing.nx_pylab.draw_networkx_edges.html}

\bibitem{Tier1Capital} Tier One Capital, \url{http://lexicon.ft.com/Term?term=tier-one-capital}

\bibitem{YahooFinance} Yahoo!\,Finance, \url{http://finance.yahoo.com/}

\bibitem{SPDR} The composition of ETFs for S\&P 500 indices
\url{http://www.sectorspdr.com/sectorspdr/}

\bibitem{SP500_ETF_list} S\&P 500 Index ETF List, \url{http://etfdb.com/index/sp-500-index/}

\bibitem{MantegnaStanleyBook} R.\,N. Mantegna and H.\,E. Stanley, {\it Introduction to Econophysics: Correlations and Complexity}
(Cambridge University Press, Cambridge, U.K., 1999).

\bibitem{WeightShift} The transformation $W_{ij} = (1 + r_{ij})/2$ ensures that all weights are nonnegative.

\bibitem{Cucuringu2013} M. Cucuringu, V.\,D. Blondel, and P. Van Dooren, Phys. Rev. E {\bf 87}, 032803 (2013).

\bibitem{MigrationCensus} US Census Bureau, 2002, \url{http://www.census.gov/2010census/}

\bibitem{Perry2003} M.\,J. Perry, Census 2000 Special Reports, 2003, \url{http://www.census.gov/population/www/cen2000/briefs/}

\bibitem{marc-subway} C. Roth, S. M. Kang, M. Batty, and M. Barth{\'e}l{\'e}my, J. R. Soc. Interface {\bf 9}, 2540 (2012).

\bibitem{dacosta} L. da F. Costa, F. A. Rodriguez, G. Travieso, and P. R. Villas Boas, Adv. Phys. {\bf 56}, 167 (2007).

\bibitem{Kolb1985} H.\,H. Kolb, J. Zool. London (A) {\bf 206}, 253 (1985).

\bibitem{White2005} C.\,R. White, J. Zool. London {\bf 265}, 395 (2005).

\bibitem{rabbit_warren_excavation_details} The rabbit warren was excavated for the purpose of filming a documentary that aired on the BBC~\cite{BBC_documentary}. The injection phase was 22--24 January 2013. The excavation phase started on 8--10 April with mechanical excavation.  A mixture of mechanical and hand excavation was done on 15--17 April.  There was exclusively hand excavation on 22--23 April, and finishing touches were applied on 30 April 2013 (while the documentary was being filmed).

\bibitem{BBC_documentary} {\it The Burrowers: Animal Underground}, \url{http://www.bbc.co.uk/programmes/b038p45r}

\bibitem{RabbitWarrenData} The simplified rabbit warren data that we used in this paper is available at
\url{https://sites.google.com/site/lshlj82/rabbit_warren_data.zip}.
There are two files: one has the node information, and the other has the edge information.

\bibitem{SimonBuckley} S. Buckley (private communication).

\bibitem{DavidHosken} D. Hosken, excerpt from the third episode of~\cite{BBC_documentary}.

\bibitem{RoadNetworkData} The road network data set is available at
\url{https://sites.google.com/site/lshlj82/road_data_2km.zip}.  The file names give the city identities.

\bibitem{MRF_code} The code to produce MRFs can be found at \url{http://www.jponnela.com/web_documents/mrf_code.zip}

\bibitem{marc-spatial} M. Barthelemy, Phys. Reps. {\bf 499}, 1 (2011).

\bibitem{SHLee2013} S.\,H. Lee and P. Holme, Eur. Phys. J. Spec. Top. {\bf 215}, 135 (2013).

\bibitem{KruskalsAlgorithm} J.\,B. Kruskal, Proc. Amer. Math. Soc. {\bf 7}, 48 (1956).

\bibitem{GoldmanBook} R. Goldman, in {\it Graphics Gems}, edited by A.\,S. Glassner (Academic, Waltham, MA, 1993), p. 304.

\bibitem{GellertBook} W. Gellert, S. Gottwald, M. Hellwich, H. K{\"a}stner, and H. K{\"u}nstner, {\it VNR Concise Encyclopedia of Mathematics} (Van Nostrand Reinhold, New York, 1989).

\end{thebibliography}
\end{document}